# Using Modified Newtonian Dynamics to Calculate the Anomalous Acceleration of Oumuamua and Partially Predict the Additional Acceleration of 3I/ATLAS

## Author

Dongcheng(Caihui) Zhao The Orphan school of Jilin 2001, China

*Orcid:0000-0001-6141-757x*

## Abstract

---

For any interstellar celestial body whose trajectory can be calculated using Newton's gravitational equation, there corresponds a unique trajectory corrected by kinematic gravitational effect. The parameters for calculating this corrected trajectory are derived from its simulated trajectory, substituting these parameters into the Modified Newtonian Dynamics equations will yield the uniquely corresponding corrected trajectory. Here using Modified Newtonian Dynamics (MOND) extended by Kinematic gravitational effect (KGE) calculated the anomalous acceleration of 1I/2017 U1'Oumuamua and made a partial prediction for the additional acceleration of 3I/ATLAS. Under the premise of excluding the gravitational influence of the planets in the solar system, during the period from Oct 19, 2017 to Jan 2, 2018, relative to the predicted trajectory position solely by the Sun's gravity, kinematic gravitational acceleration will lead to an extra trajectory deviation about *40,732 km* in Oumuamua's trajectory, close to the *40,000 km* given by astronomical observations. On Oct 25, 2017 12:00 at 1.36 AU the comparable acceleration caused by KGE is $2.68\times10^{-6}\ m/s^2$, which is very close to the value in orbital fits in 2018 that $A1=2.7 \times 10^{-6}\ m/s^2$ the margin of error between them is less than 2%.

Similarly, Atacama Large Millimeter Array (ALMA) indicated that 3I/ATLAS was *4 arcseconds* away in Right Ascension from where it was supposed to be on October 29, 2025 if its trajectory was dictated by gravity. At perihelion, the observed non-gravitational acceleration $A_1$ is about $134.26\ km/day^2$ radially, $A_2$ is about $57.35\ km/day^2$ transversely. This equals an acceleration $146.00\ km/day^2$ or $0.0196\ millimeter/s^2$ at moving direction, which is close to the MOND predicted comparable accelerations $Ac_{1.36}$ and $Akc_{1.36}$ of $139.97\ km/day^2$ or $0.01875\ millimeter/s^2$ at moving direction, the margin of error is near *4%*. When 3I/ATLAS arriving at perigee on Dec 19 2025, from Jul 1, 2025 to Dec 19, 2025, relative to the predicted position solely by the Sun's gravity, the kinematic gravitational acceleration will make it travel an additional distance approximately *19,134 km* in its trajectory.

## Main

On October 19, 2017, Pan-STARRS1 astronomical telescope discovered an asteroid with relatively high velocity and orbital eccentricity of *1.92*, named 1I/2017 U1'Oumuamua. It was our first known interstellar visitor. Under the premise of excluding the gravitational influence of the planets in the solar system, scientists have found small discrepancies between its observed trajectory and the trajectory dictated by gravity, until Jan 2, 2018, the observed results was *40,000 kilometers* longer than the predicted results,[1] and until May 3, 2018, the difference should be *100,000 kilometers*.[2,3] This means that there is a significant anomalous acceleration in is trajectory,[4,5] which is currently mostly thought to be caused by some non-gravitational reasons, such as cometary outgassing.[6,7,8,9] However, based on the observations of Spitzer Space Telescope, since no obvious evidence of outgassing has been observed,[10,11,12,13] it has been classified as an asteroid, so its anomalous acceleration is difficult to be explained by cometary outgassing theories. There are still some other non-gravitational explanations,[14,15,16,17,18] but none of them have been verified. Due to Oumuamua has left the solar system and will never return, so it is impossible to verify them by further observation.

However, there is a new explanation of the anomalous acceleration of Oumuamua based on Modified Newtonian Dynamics, which can be verified indirectly. If the anomalous accelerations are caused by a hidden gravitational factor that hasn't described by Newton's gravitation equation, then the explanation should not only be able to calculate the anomalous acceleration of Oumuamua, but also predict the additional acceleration of newly discovered interstellar asteroids or partially predict the additional acceleration of newly discovered interstellar comets. In case the predict results consistent with the astronomical observations, it equals an indirect verification of the anomalous acceleration of Oumuamua.

Newtonian gravity is assumed to be an instantaneous force, which determines the gravitational effect between objects is independently of their relative velocity. However, if gravity has kinematic properties, it should take time to travel, then the kinematic gravitational effect between objects will be related to the square of their relative velocity, this means that the kinematic gravitational effect between objects will change as the square of their relative velocity changes. Under the premise that gravity travels at the speed of light, a new Modified Newtonian Dynamics is proposed here to describe the anomalous acceleration of Oumuamua, and it can also be used to make a partially prediction for the additional acceleration of 3I/ATLAS.

The kinematic gravitational effect assumes gravity travels at speed of light, in this case gravity not only depends on mass of objects and square of distance between them, but also depends on the relative velocity and $\alpha$ angle between them.

The Newton's gravitational equation is:

$$F = G\frac{Mm}{r^2} \tag{1}$$

Therefore, based on the theory of Kinematic gravitational effect, the relationship between Modified Newtonian Dynamics and Newton's law of gravity can be described by using geometric method in the *fig1*.

*Figure 1*

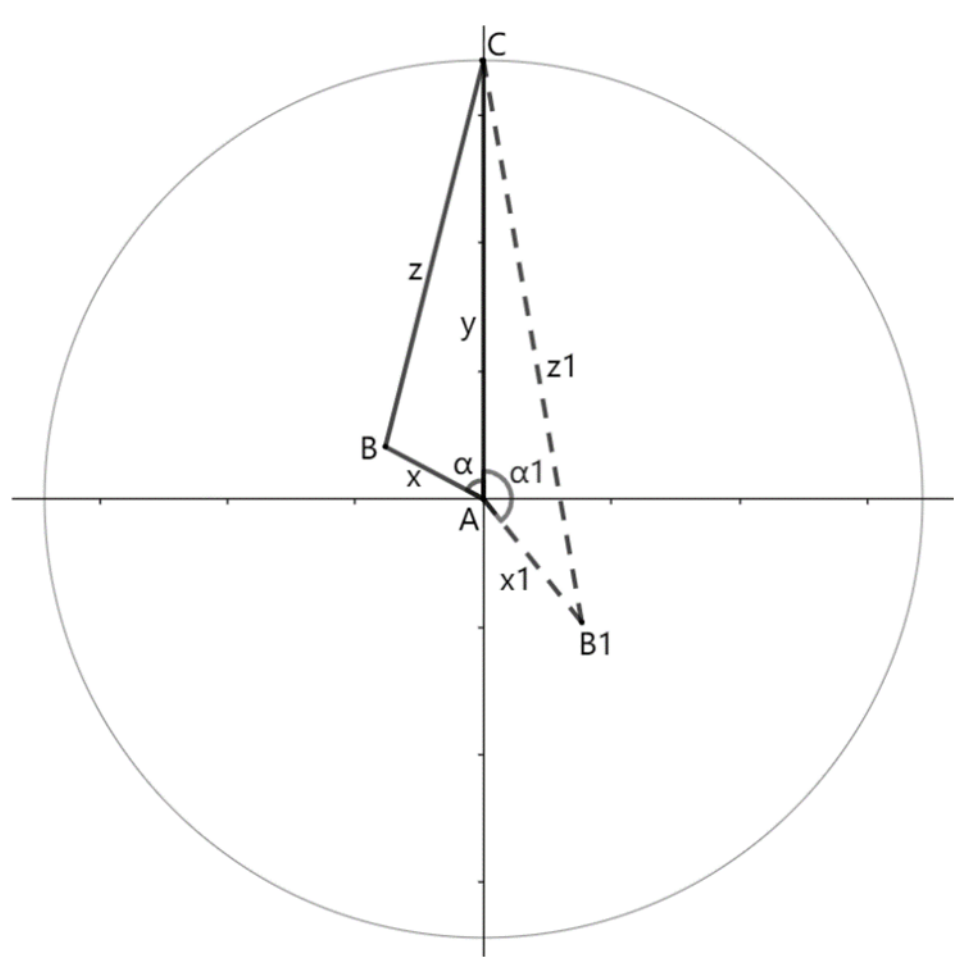

In a gravitational field,

$x$ represents the object moving from direction A to B, at velocity of $v$.

$y$ represents the gravity travels from direction A to C, at velocity of light (*299792.458 km/s*).

$\alpha$ = the included angle between $x$ and $y$, in gravitational field, it is shown as the angle between the extension of the line connecting the gravitational source and the object with the moving direction of the object. In applications, it is a three-dimensional spatial angle, when the viewpoint is perpendicular to the orbital plane of an object, it can be approximated as a planar angle for measurement.

$F$ is the Newton's gravitational force on static object, its strength can be represented by the length of $y$.

$F_k$ is the gravitational force on moving object which incorporates the Newton's gravitational force on static object and the kinematic gravitational effect on moving object, its strength can be represented by the length of $z$. Because z does not have a velocity component, so when the value of $z$ is greater than $y$, it does not mean its velocity is greater than the speed of c, but just means the object is under greater gravitational force than that it is at static.

In *figure 1*, $x$, $y$ and $z$ formed a triangle $\Delta ABC$, then the relationship between them can be solved with the Cosine theorem, so the relationship between $F_k$ and $F$ can be described by the following equation:

$$F_k = \frac{\sqrt{v^2 + c^2 - 2vc\cos\alpha}}{c} F \qquad (2)$$

Thus, the form of the Newton's gravitation equation while incorporating the kinematic gravitational effect, can be expressed as:

$$F_k = G\frac{\sqrt{v^2 + c^2 - 2vc\cos\alpha}.Mm}{cr^2} \qquad (3)$$

Relative to the gravitational field, when an object remains static, its relative velocity *v=0,* then equation (2) and (3) = equation (1), the Newton's gravitation equation remains unchanged, then $F_k=F$; when $v>0$, then $F_k - F$ is the value of kinematic gravitational effect.

The kinematic gravitational effect will also be affected by the α angle, when a celestial body's orbital eccentricity is close to *0,* the corresponding α angle is close to *90*°, then the effect is weakest. If the orbital eccentricity of a celestial body is close to 0 and its velocity is far less than the speed of light, it is difficult to directly observe the influence of this effect on the object's orbit.

For an example, the magnifications of gravitational force of a celestial body with the velocity of *30 km/s* in different α angle are shown in ***Table 1***:

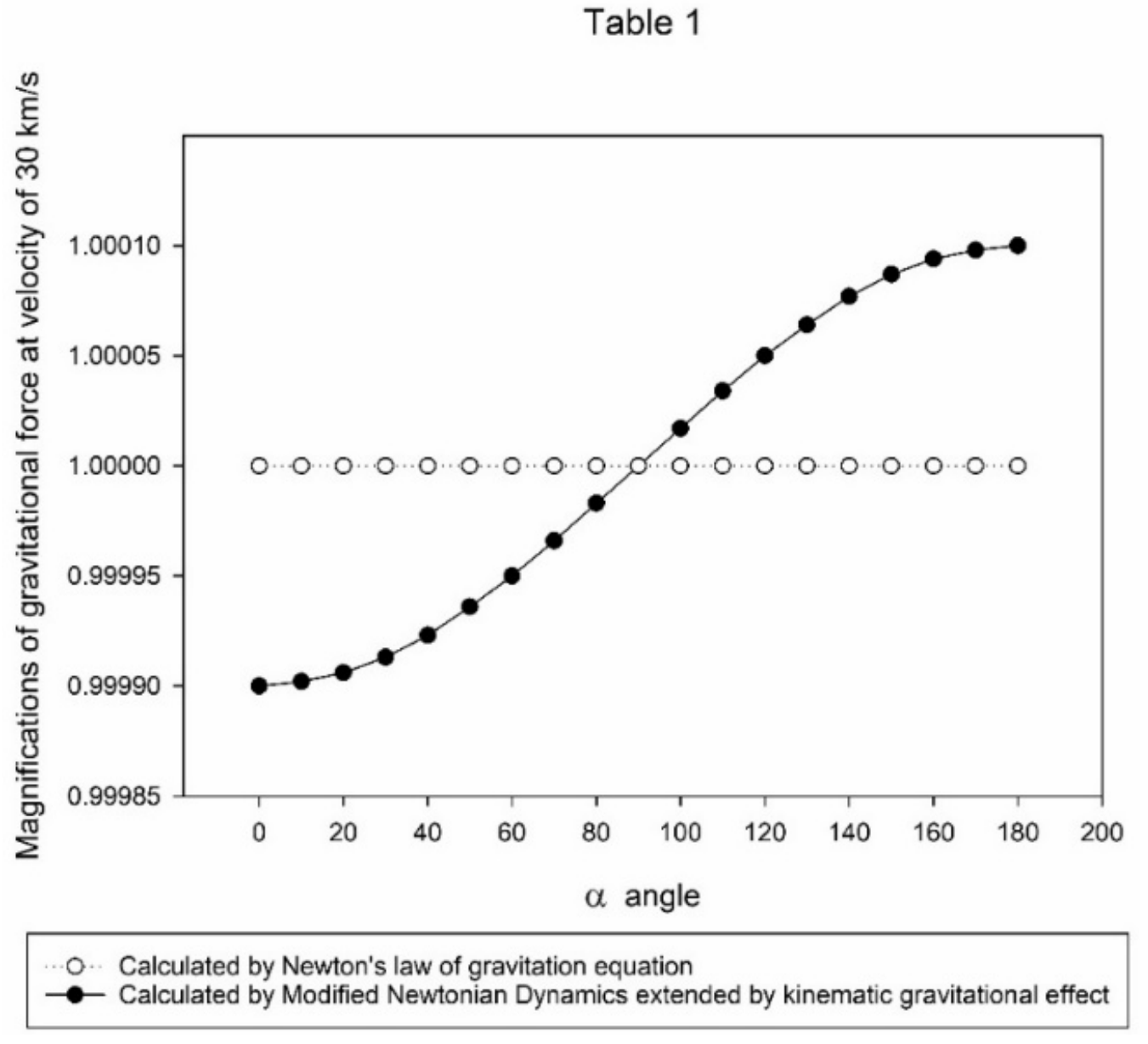

The eccentricity of the Earth's orbit is 0.0167086, for most of the year, its α angle near *90°*, so relative to the result calculated by Newton's gravitational equation, the instant difference caused by Modified Newtonian Dynamics may less than one millionth which is hardly to observe directly. However, as the increase of some celestial bodies's moving speed and the increase of their orbital eccentricities, the difference between the calculated results of equation (1) and equation (2) (3) will also increase, and making it easier to observe.

Oumuamua moving at a relative high velocity in the solar system, and its orbital eccentricity is *1.92*, so it should be significantly influenced by kinematic gravitational effect. In case the anomalous acceleration of Oumuamua is generated by kinematic gravitational effect, then it should able to be described by equation (2) or equation (3). By bring the orbital parameters of Oumuamua in to the modified equations, then it is able to know whether the calculated results agree with the astronomical observed results.

## Methods

---

**The calculated trajectory deviations of Oumuamua by using Modified Newton's gravitation equation (2), (3) during Oct 19, 2017 – May 3, 2018**

Based on the theory of the kinematic gravitational effect, Oumuamua will get additional accelerations over time during its whole journey in the solar system, due to it had already past the perihelion before it was discovered, so there is no way to confirm whether the anomalous acceleration was existed before the date it has been discovered by directly astronomical observe. However, the kinematic gravitational acceleration would be accumulated into velocity increments over time, and it would not disappear without interference of external forces, after pass perihelion, the Sun's gravity would decelerate Oumuamua's main velocity, and the velocity increments caused by KGE would decelerate at the same proportion. Relative to the predicted position solely by gravity, the velocity increments would lead to an additional distance in its trajectory, and this can be observed after a sufficiently long period of time. Under the premise the gravity influence of planets in solar system are excluded, according to the observational data, by compare the observed trajectory with the predicted trajectory position dictated by gravity, (since Oct 19, 2017) location of Oumuamua had been boosted for about *40,000 km* until Jan 2, 2018,[1] and calculated to be *100,000 km* until May 3, 2018.[2,3]

Relative to Oumuamua's velocity $v$, kinematic gravitational acceleration will accumulate to be different terminal velocities $v_k$ over different time periods, with $v_k - v$ is the velocity increment. Due to $F$ and $F_k$ are the reason the $v$ and $v_k$ increase or decrease, so:

$$\frac{F_k - F}{F} = \frac{v_k - v}{v} \tag{4}$$

then,

$$v_k - v = \frac{F_k - F}{F} v \tag{5}$$

Oumuamua's trajectory data for calculation is cited from CSS (Catalina Sky Survey) Orbit Viewer (updated 7/19/2025) which is based at the University of Arizona's Lunar and Planetary Lab,[19] using data from the International Astronomical Union Minor Planet Center (MPC),[20,21] made trajectory simulations of Oumuamua and 3I/ATALS in solar system, their trajectories information can be seen in these simulation, such as velocity, α angle and distance from the Sun at each moment. (The Oumuamua's trajectory simulation in old version is cited from Spacein3D)[22] Since α angles are not shown in simulation, thus they are measured manually. (Objects plot from Minor Planet Center, 1I for Oumuamua, 3I for 3I/ATLAS. In order to ensure the viewpoint is perpendicular to trajectory plane of Oumuamua, click the menu located in the upper-left corner inside the box, and adjust it to perihelion by reschedule to around 10:30 PM on September 9, 2017, then through click and drag, zoom in and out, make sure Oumuamua lies on the central longitudinal axis passing through the Sun, align the inbound trajectory with the top-left corner of the box and the outbound trajectory with the top-right corner of the box in the screen, then the perpendicular view of the hyperbolic trajectory of Oumuamua can be got as shown in *fig 2*. In this case, α angles can be obtained by directly measure. )

***Figure 2***

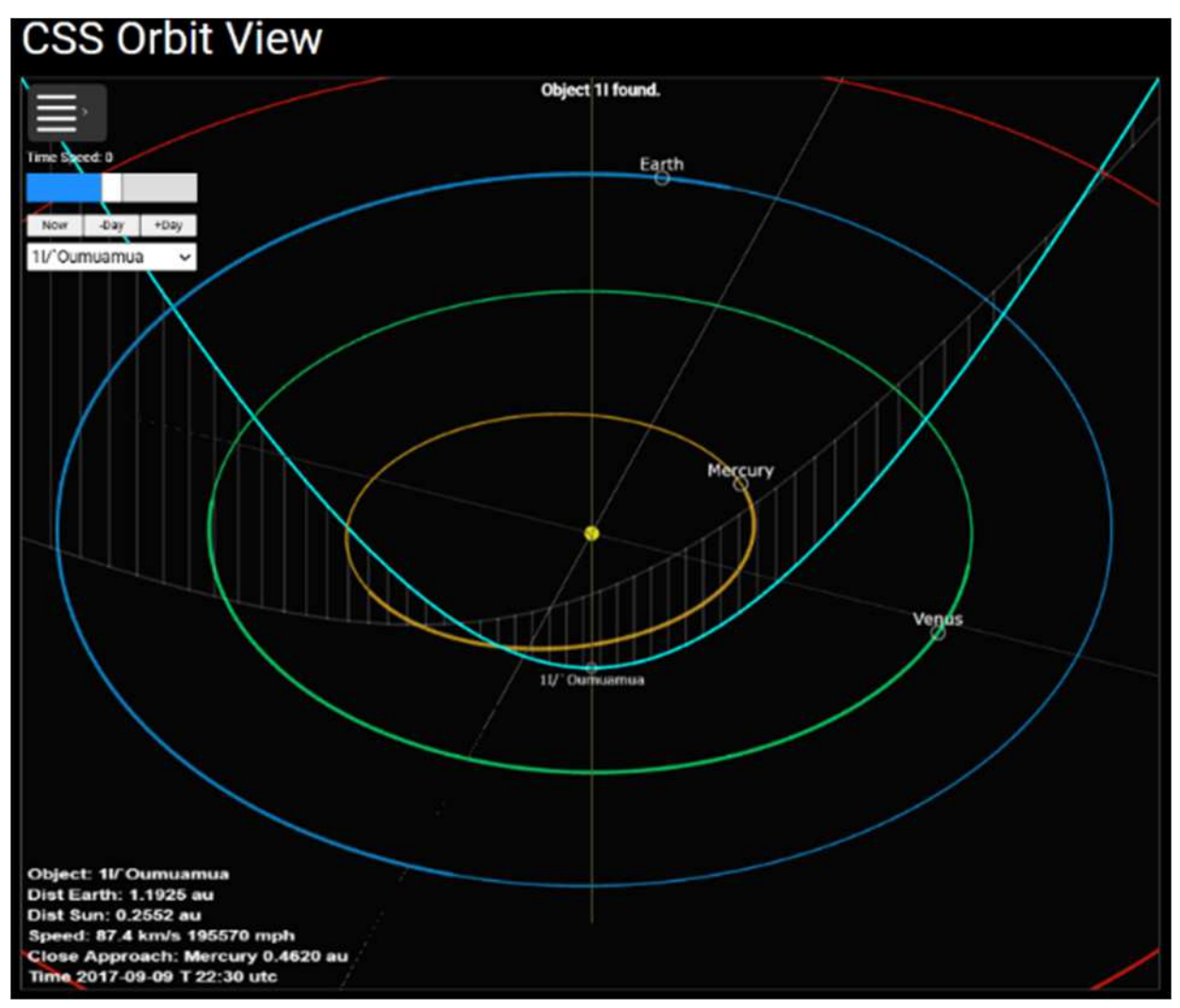

Bringing these data in to equation (2), (3), it is possible to calculate the instantaneous value of kinematic gravitational effect on these interstellar objects, then with the help of their velocity change, it is able to calculate the velocity increment accumulated by kinematic gravitational acceleration on their trajectories. After calculation (the calculation process is at the calculation section which is under the conclusion section), curve of velocity increments accumulated by kinematic gravitational acceleration of the Sun on different dates of Oumuamua's trajectory is shown in ***Table 2***.

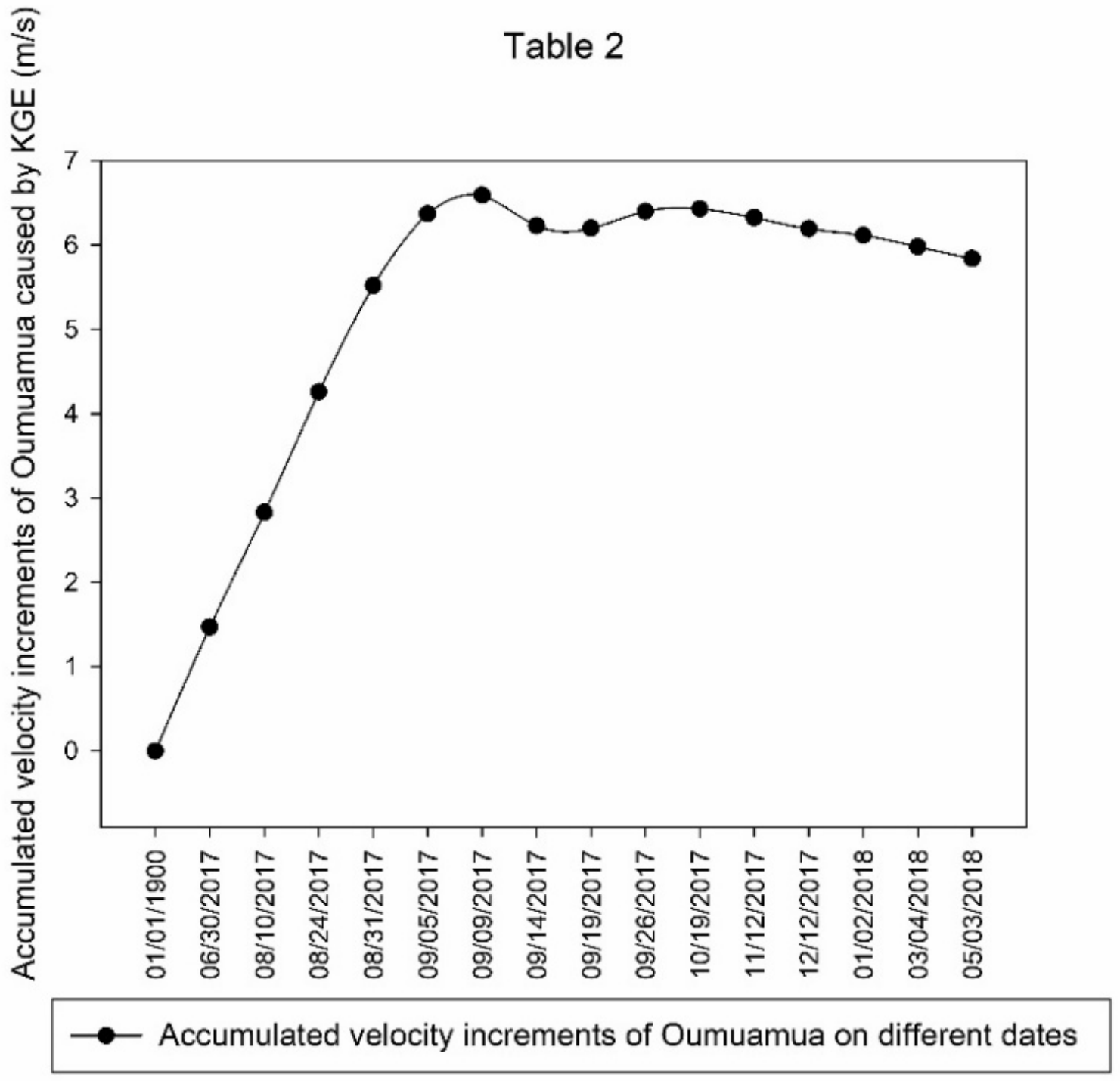

Due to varying time intervals between adjacent cells on the X-axis, the X-axis cells in tables should be expanded alone the timeline to obtain a velocity increment curve approximating the actual time flow rate.

In order to compare the velocity increments in ***table 2*** with the orbital fits (Micheli et al) or some observed non-gravitational acceleration, two comparable acceleration $A_c$ and $A_{kc}$ can be used for comparison, $A_c$ follows the non-gravitational acceleration rule, but $A_{kc}$ follows kinematic gravitational acceleration rule, on perihelion $A_c = A_{kc}$.

Based on acceleration formula:

$$\boldsymbol{a} = \frac{\boldsymbol{vt - v0}}{\boldsymbol{t}} \tag{6}$$

$A_c$ and $A_{kc}$ can be obtained by accumulated velocity increments divide the accelerating time of 86400 seconds. When Oumuamua passing the perihelion on Sep 9 2017 at 0.255 AU to the Sun, when $v_0=0$, the results calculated by MOND show that the velocity increment $6.59\ m/s$ will lead to two comparable acceleration $A_c$ and $A_{kc}$ with the value of $7.63\times10^{-5}\ m/s^2$ or $3.81\times10^{-6}\ AU/day^2$ at moving direction, which can be composed by radial acceleration $A1$ and transverse acceleration $A2$, with the relationship $A = \sqrt{A1^2 + A2^2}$. $A_c$ will attenuated to be $4.96\times10^{-6}\ m/s^2$ at the distance 1 AU from the Sun, the value in the orbital fits at 1 AU is $(4.92 \pm 0.16)\times10^{-6} ms^{2}$, and at the distance 1.36 AU (approximately 1.4 AU) on Oct 25, 2017 12:00 AM, $A_c$ will attenuate to be $A_{c1.36}=4.96\times10^{-6}/1.36^2=2.68\times10^{-6}\ m/s^2$, the value in the orbital fits on October 25 given by (Micheli et al) is $A1=2.7\times\ 10^{-6}\ m/s^2$, and at 1.4 AU it will attenuate to be: $A_{c1.40}=4.96\times10^{-6}/1.40^2= 2.53\times10^{-6}\ m/s^2 =2.53\times10^{-4}\ cm/s^2$, close to the value given by (Seligman et al) of $2.5\ \times\ 10^{-4}\ cm/s^{-2}$. The orbital fits assumed $A2$ close to $0$,[4] so the results calculated by MOND and the acceleration in the orbital fits are very close to each other, the margin of error between them is less than $2\%$.

$A_{kc}$ is a hypothetical acceleration at moving direction that can compare with non-gravitational acceleration only of perihelion, in order to illustrate the magnitude of the acceleration resulting from the cumulative velocity increment at a specified date relative to the non-gravitational acceleration at perihelion. $A_{kc}$ assumes that all velocity increments accumulated prior to the specified date occurred on that date, it does not follow the inverse square law; its magnitude is determined by the accumulated velocity increments at the specified date. The velocity increment of an object increasing by accumulate kinematic gravitational acceleration over time, its curve is always be ascending curve before perihelion, after perihelion, while the velocity increment increases with the accumulation of kinematic gravitational acceleration, it also decreases proportionally with the decrease of the its primary velocity. MOND calculated that Oumuamua's accumulated velocity increment on Oct 19 2017 at 1.20 AU is *6.43 m/s*, until Jan 2 2018 at 2.86 AU it only decreased to be *6.14 m/s*, the velocity increment will result in a comparable acceleration $A_{kc}$ which remain almost unchanged respect to the inverse square law, should be the strongly disfavored constant acceleration described by (Micheli et al) that independent of distance, regardless direction.

Through calculations, compare with the predicted position solely by gravity, during the observation period of Oct 19, 2017 – Jan 2, 2018, Oumuamua traveled an additional *40,732 km* under the influence of KGE, close to the observation result of *40,000 km*, with the margin of error less than *2%*. The observation records from the Minor Planet Center (MPC) indicate that the last recorded observation of Oumuamua was made by the Hubble Space Telescope on Jan 2, 2018, subsequently, it became impossible to continue observing it due to a significant decrease in its brightness. Two dates, Nov 12, 2017 and Dec 12, 2017 were selected from the MPC's observation records, during the period from Oct 19 2017, relative to the predicted trajectory position solely by gravity, calculations indicate that Oumuamua should have moved an additional *13,250 km* and *29,528 km* on the corresponding dates. Theoretically, it is possible to calculate the trajectory deviations of Oumuamua on any given days within the observation period, and then compare them with the observed results for further verifications.

## The Kinematic gravitational acceleration of 3I/ATLAS

---

On July 1, 2025, the NASA-funded ATLAS (Asteroid Terrestrial-impact Last Alert System) survey telescope discovered a comet with high velocity and orbital eccentricity near 6.14, named 3I/ATLAS. Based on the theory of Modified Newtonian Dynamics, 3I/ATLAS should be accelerated by Kinematic gravitational effect. CSS Orbit Viewer (update 08/05/2025) using data from the MPC,[19,21] established the trajectory simulation of 3I/ATALS at the early observation stage, by using the above method, it is able to calculate the influence of acceleration caused by KGE on the future trajectory of 3I/ATALS. After calculations, curve of instantaneous kinematic gravitational acceleration $A_k$ at moving direction is shown in ***Table 3,*** curve of accumulated velocity increments caused by kinematic acceleration on different days is shown in ***Table 3.1***. During the period from the date Jul 1 2025, relative to the predicted trajectory position solely by gravity, curve of its trajectory deviations caused by kinematic acceleration on different dates is shown in ***Table 4***.

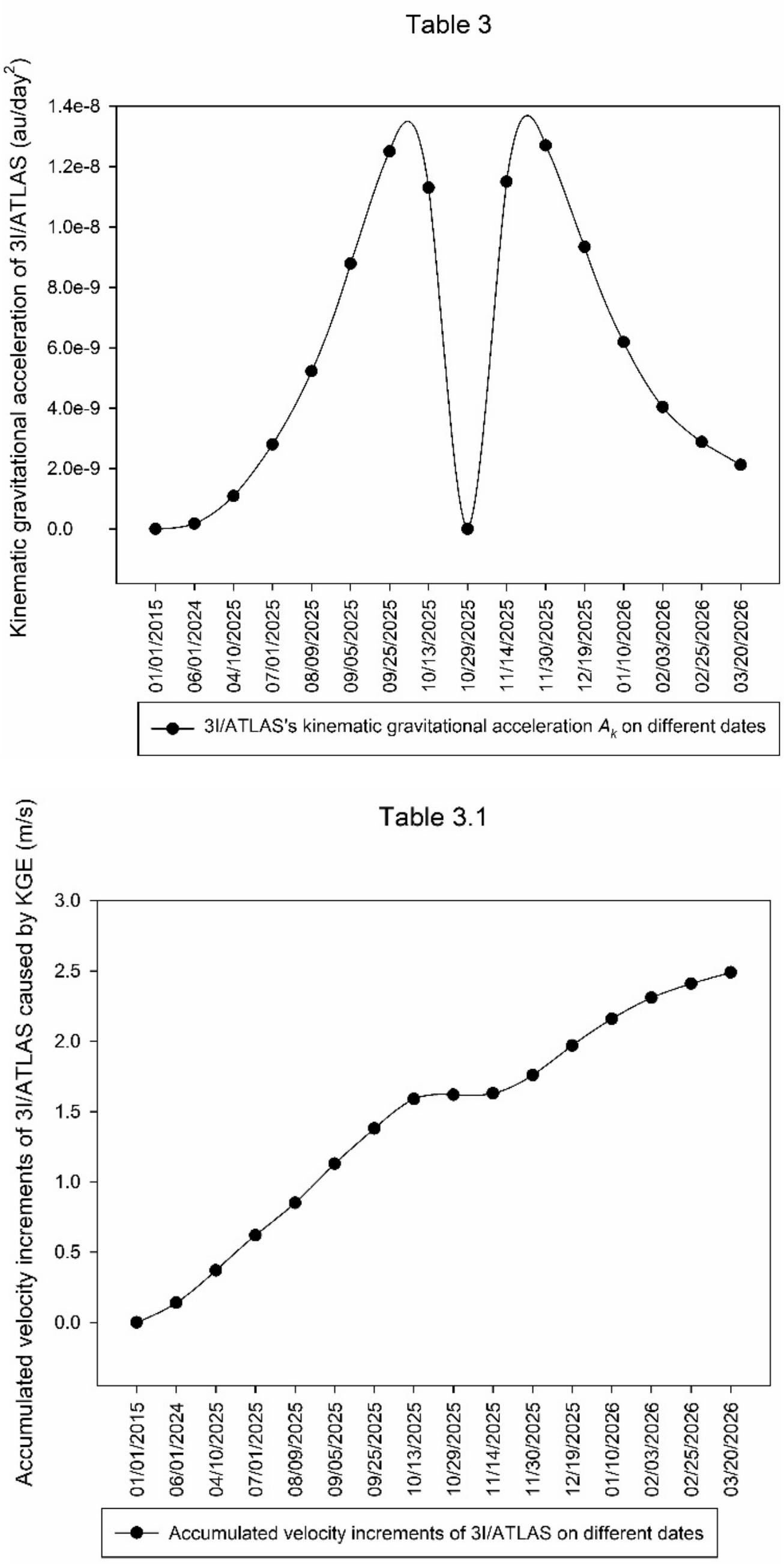

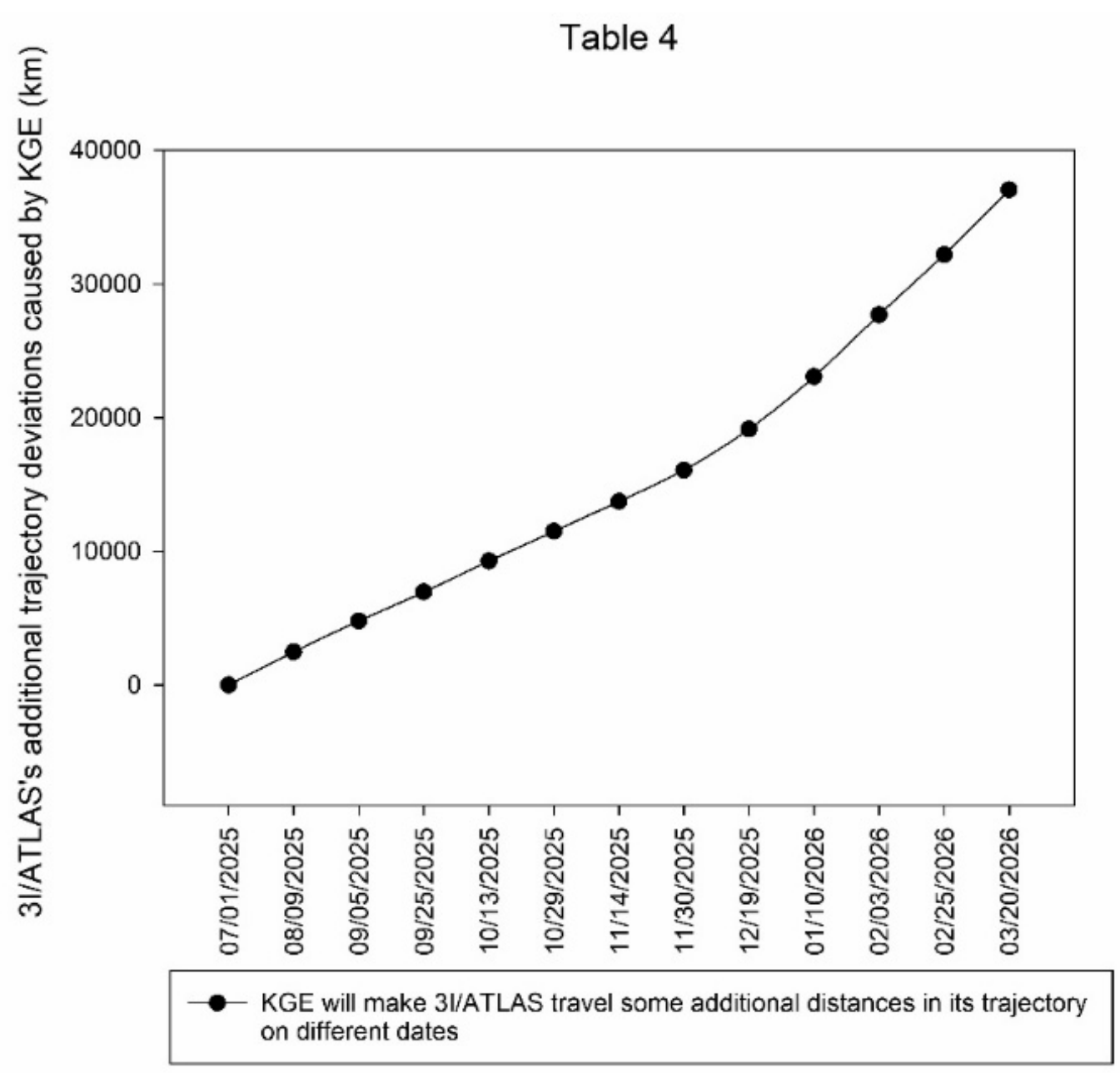

Without considering the influence of gravity of planets in solar system, since 3I/ATLAS has been confirmed as a comet,[23,24,25,26,27,28] therefore the curves shown in *Table 3.1, Table 3.2* and *Table 4* must incorporate the influence of cometary outgassing to match the astronomical observations.

Atacama Large Millimeter Arry (ALMA) indicate that 3I/ATLAS was *4 arcseconds* away in Right Ascension from where it was supposed to be on October 29, 2025 if its trajectory was dictated by (static) gravity. Based the observation and some earlier observations a non-gravitational acceleration report is given by Davide Farnocchia that about $A_1=1.66\times10^{-6}\ AU/day^2$ radially, $A_2=7.09\times10^{-7}\ AU/day^2$ transversely at 1 AU.[29] At perihelion (1.36 AU), when without consider the uncertainty, $A1_{1.36}$ should be $1.66\times10^{-6}/1.36^2=8.97\times10^{-7}AU/day^2$ or $134.26\ km/day^{-2}$ radially, $A2_{1.36}$ should be $7.09\times10^{-7}/1.36^2=3.83\times10^{-7}AU/day^2$ or $57.35\ km/day^{-2}$ transversely, equal an acceleration of $A_{1.36}$ at moving direction that:

$$A_{1.36} = \sqrt{A1^2 + A2^2} = \sqrt{134.26^2 + 57.35^2} = 146.00\ km/day^{-2} = 9.76\times10^{-7}\ AU/day^2 = 1.96\times10^{-5}\ m/s^2$$

The conversion ratios between different acceleration units are approximately:

$1\ m/s^2 = [1 / 1.496\times10^{11}] \times [86400^2]\ AU/day^2 = 0.0499\ AU/day^2$

$1\ AU/day^2 = 1.496\times10^8\ km/day^2$

The non-gravitational acceleration of 3I/ATLAS at perihelion gave by Avi-Loeb is $A1_p$=*135 km/day*$^{-2}$ radially, $A2_p$=*60 km/day*$^{-2}$ transversely, $A_p$=*0.02 millimeter/s*$^2$. [30,31]
At perihelion, by using equation (6), predicted comparable accelerations $A_c$ and $A_{kc}$ at moving direction based on the velocity increments *1.62 m/s* caused by KGE should be: $A_{c1.36}=A_{kc\ 1.36}=1.62/86400=1.875\times10^{-5}\ m/s^2\ =9.36\times10^{-7}\ AU/day^2\ =139.97\ km/day^2$. The predicted kinematic acceleration value *139.97 km/day*$^2$ is close to the observed value *146.00 km/day*$^{-2}$, *(146.00 -139.97)/ 146.00=0.041,* the margin of error is close to *4%*, the errors may cause by cometary outgassing and some other reasons. All of the curves in *Table 3.1, Table 3.2* and *Table 4* are based on the kinematic gravitational acceleration data in *Table 3* and in the calculation processes that follow the conclusion section, all data parameters used in the calculation are non-adjustable and determined by the simulated trajectory of 3I/ATLAS.[19] When the trajectory parameters of an interstellar object are input into the equation of MOND, only one unique set of computational results will be output. The final trajectory deviation results for comparison with observations in *Table 4* was included in the previous *v14* version submitted on Oct 27 2025,[32] which is earlier than ALMA and Davide Farnocchia indicated the non-gravitational acceleration on Oct 29 2025.[29]

When calculating the corrected trajectory of interstellar bodies, to reduce computational complexity, representative time points are typically selected at different stages. By substituting the corresponding trajectory parameters of these time points into the equation of MOND, an approximate calculation of the corrected trajectory can be obtained. The selection of time points may affect the accuracy of the computational results, once these time points are specified, the computational results will be unique. Since the simulated trajectory calculated using Newton's gravitational equation uniquely corresponds to a corrected trajectory computed via MOND equations, even though the previous *v14* version does not include the computational process and corresponding velocity increment table, the velocity increment data at specific time points can still be inversely derived from the final computational results.

Furthermore, on Oct 27 2025, the full version of the manuscript which included the computational process and corresponding velocity increments table has been sent to an online storage server by trusted official uploader, the manuscript can be downloaded from the online address.[33] After download, the author information can be viewed via Microsoft Word or through Right-click the document file, select Properties, then Details to display. The address of the online file contains the complete creation date information; this is evidence that the velocity increment of *1.62 m/s* at perihelion calculated by MOND is a predicted value.

Comet outgassing, the accuracy and time span of simulated trajectory, the selection of time points, the precision of α angle, and the accuracy of computational values, all these factors can affect the precision of the calculation results.

## Discussions

---

The variation in the intensity of non-gravitational acceleration produced by cometary outgassing follows the inverse square law of the distance from the Sun, after perihelion, non-gravitational acceleration of celestial bodies will manifest as a descending curve. After perihelion, 3I/ATLAS's velocity increments continue to accumulate under the influence of kinematic gravitational acceleration while decreasing proportionally with the reduction in its primary velocity. Since the cumulative efficiency of its velocity increments exceeds the efficiency of its primary velocity decrease, therefor its velocity increments will manifest as an ascending curve. In subsequent reports released by Davide Farnocchia, an anomalous increase in non-gravitational acceleration due to $CO_2$ was observed as 3I/ATLAS moved away from the Sun. This phenomenon cannot be explained by cometary outgassing but aligns with MOND's predictions. [34,35,36]

Similarly, it is able to partially predict the additional acceleration of a new interstellar objects by using MOND, but there are no new interstellar objects have been discovered at present. However, there is a new comet named C/2025 R3 (PanSTARRS) will arrive at perihelion on April 20, 2026, with the eccentricity of 1.0003, this indicates its orbit will be a hyperbolic trajectory. CSS Orbit Viewer has established its trajectory simulation,[19] type "2025 R3" will plot it from MPC. By substituting the parameters from trajectory simulation in to equation (2), I calculated its additional acceleration at moving direction caused by KGE of the Sun. At its perihelion (0.499 au), the cumulative velocity increments $vt$ is near $4.03\ m/s$, the corresponding comparable acceleration $A_c$ and $A_{kc}$ at moving direction should be:

$A_{c0.499}=A_{kc0.499}=4.03/86400=4.66\times10^{-5}\ m/s^2=2.33\times10^{-6}\ AU/day^2=348.20\ km/day^2$.

$A_{c1.0}$ should be: $A_{c0.499}\times0.499^2/1^2=1.16\times10^{-5}\ m/s^2=5.80\times10^{-7}\ AU/day^2=86.70\ km/day^2$.

Errors may cause by the reasons described earlier.

## Conclusion

The calculated result of the anomalous acceleration of Oumuamua using the Newtonian gravitation equation extended by kinematic gravitational effect essentially consistent with the astronomical observations during the period of Oct 19,2017- Jan 2, 2018 , the margin of error is less than *2%*. For the additional acceleration of 3I/ATLAS at perihelion, when without consider the uncertainty, the margin of error between the predicted value calculated by Modified Newtonian Dynamics and the astronomical observed value is only about *4%*, this indicates that MOND is worthy of serious consideration and deeply study. Under the premise that without consider the influence of gravity of planets in solar system, when 3I/ATLAS reaches its perigee on Dec 19 2025, relative to the predicted trajectory position solely by the Sun's gravity, from Jul 1 2025 to Dec 19 2025, the kinematic gravitational acceleration will make 3I/ATLAS move an additional *19,134* kilometers in its trajectory, and after it passed Jupiter's orbit on Mar 20, 2026, the additional distance is going to be *37,048* kilometers. Theoretically, this method should be able to predict the additional acceleration of newly discovered interstellar asteroids or partially predict the additional acceleration of newly discovered interstellar comets, when their trajectory simulation established.

## Calculations

### Specific calculation process of Oumuamua's accumulated velocity increments and trajectory deviations caused by KGE

#### The calculation during the period of inbound the solar system

Oumuamua was already on its way out of the solar system when it was discovered on Oct 19 2017, and scientists have calculated that it entered solar system at an initial speed of roughly *26,400 m/s* a long time ago, the calculation starts at this speed.

Based on trajectory data of Oumuamua the calculations of equation (2) show that:

*When* $v=26{,}400\ m/s$, $\alpha=179^{o}$ *then* $F_k=(1+8.8\times10^{-5})F$ *Jan 1, 1900*

*When* $v=40{,}000\ m/s$, $\alpha=163^{o}$ *then* $F_k=(1+1.28\times10^{-4})F$ *Jun 30, 2017*

Bring these values in to equation (5), the calculation shows that:

$$v_k - v =(8.8\times10^{-5}+1.28\times10^{-4})/2\times(40{,}000-26{,}400)=1.47\ m/s$$

This calculation means that Oumuamua will get a velocity increment of *1.47m/s* caused by kinematic gravitational effect, during its velocity increased to *40,000m/s* from *26,400m/s*.

Similarly, it can be inferred that:

*When* $v=50{,}000\ m/s$, $\alpha=150^{o}$ *then* $F_k=(1+1.44\times10^{-4})F$ *Aug 10, 2017*

*Interval* $v_k - v =(1.28\times10^{-4}+1.44\times10^{-4})/2\times(50{,}000-40{,}000)=1.36\ m/s$

At the end of this stage Oumuamua's velocity increment should accumulated to be:

*Cumulative* $vt=1.47+1.36=2.83\ m/s$

Similarly,

*When* $v=60{,}000m/s$, $\alpha=135^{o}$ *then* $F_k=(1+1.42\times10^{-4})F$ *Aug 24, 2017*

$v_k - v =(1.44\times10^{-4}+1.42\times10^{-4})/2\times(60{,}000-50{,}000)=1.43\ m/s$

$vt=2.83+1.43=4.26\ m/s$

*When* $v=70{,}000\ m/s$, $\alpha=118^{o}$ *then* $F_k=(1+1.10\times10^{-4})F$ *Aug 31, 2017*

$v_k - v =(1.42\times10^{-4}+1.10\times10^{-4})/2\times(70{,}000-60{,}000)=1.26\ m/s$

$vt=4.26+1.26=5.52\ m/s$

*When* $v=80{,}000\ m/s$, $\alpha=103^{o}$ *then* $F_k=(1+6.01\times10^{-5})F$ *Sep 5, 2017*

$v_k - v =(1.10\times10^{-4}+6.01\times10^{-5})/2\times(80{,}000-70{,}000)=0.85\ m/s$

$vt=5.52+0.85=6.37\ m/s$

*When* $v=87{,}400\ m/s$, $\alpha=90^{o}$ *then* $F_k=(1+4.25\times10^{-8})F$ *Sep 9, 2017*

$v_k - v =(6.01\times10^{-5}+4.25\times10^{-8})/2\times(87{,}400-80{,}000)=0.22\ m/s$

When Oumuamua arrived at the perihelion, the difference in its inbound velocity increment between respectively calculated by equation (2) and equation (1) will accumulate to be:

$vt=6.37+0.22=6.59\ m/s$

**The calculation during the period of outbound the solar system**

When Oumuamua moving outbound the solar system in the hyperbolic trajectory, the effect of the sun's gravity on its velocity is dominated by deceleration. During this period, the gravitational force on Oumuamua calculated by equation (2) is a bit weaker than the calculation of Newton's gravitation equation, this equivalent to it obtained an additional acceleration. Based on Oumuamua's trajectory data, the calculation of equation (2) shows that:

*When* $v=87{,}400\ m/s,\ \alpha=90^o$ *then* $F_k=(1+4.25\times10^{-8})F$ $(r_{sun}=0.255\ au)$ *Sep 9, 2017*

Typically, non-gravitational acceleration begins at the time just past perihelion, in order to compare the velocity increment caused by KGE with the non-gravitational acceleration, here should transform it into acceleration. Based on acceleration formula: $\boldsymbol{a}=\frac{\boldsymbol{vt}-\boldsymbol{v0}}{\boldsymbol{t}}$, when $v_0=0,\ v_t=6.59\ m/s,\ t=24\times3600=86400s,$ the comparable acceleration $A_c$ and $A_{kc}$ caused by KGE at perihelion should be:

$$A_{c0.255}=A_{kc0.255}=6.59/86400=7.63\times10^{-5}\ m/s^2=3.81\times10^{-6}\ AU/day^2$$

Based on MOND theory, the velocity increment of *6.59 m/s* was accumulated by kinematic gravitational acceleration from Jan 1 1900 to Sep 9 2017, the accelerating time *t* is much larger than it in non-gravitational acceleration calculation. The distance to the sun at perihelion is 0.255 au, because cometary outgassing is inversely proportional to the square of its distance from the Sun, so when it at 1 AU , $A_{c1}$ should be:

$$A_{c1}=A_{c0.255}\times0.255^2/1^2=7.63\times10^{-5}\times0.255^2/1^2=4.96\times10^{-6}\ m/s^{-2}=2.48\times10^{-7}\ AU/day^2$$

*When* $v=80{,}000\ m/s,\ \alpha=77^{\,o}$ *then* $F_k=(1-6.0\times10^{-5})F$ *Sep 14, 2017*

The velocity increment accumulated in this period is:

$$v_k-v=(4.25\times10^{-8}-6.00\times10^{-5})/2\times(80{,}000-87{,}400)=0.22\ m/s$$

This calculation means that Oumuamua will get a velocity increment of *0.22m/s* caused by kinematic gravitational effect, during the period that its velocity decrease to *80,000 m/s* from *87,400 m/s*.

After perihelion, due to the deceleration of the Sun's gravity, the velocity increment caused by KGE decreasing synchronously at same reduction ratio while Oumuamua's primary velocity decreasing, after its primary velocity decreased to *80 km/s* from *87.4 km/s,* the velocity increment should approximately to be:

$$vt= (6.59+0.22)\times 80/87.4=6.23\ m/s$$

Similarly, it can be inferred that:

When $v=70{,}000\ m/s,\ \alpha=62^o$ then $Fk=(1-1.10\times 10^{-4})F$ Sep 19, 2017

$$v_k - v=(-6.0\times 10^{-5}-1.10\times 10^{-4})/2\times(70{,}000-80{,}000)=0.85\ m/s$$

$$vt= (6.23+0.85)\times 70/80=6.20\ m/s$$

When $v=60{,}000\ m/s,\ \alpha=45^o$ then $F_k=(1-1.42\times 10^{-4})F$ Sep 26, 2017

$$v_k - v=(-1.10\times 10^{-4}-1.42\times 10^{-4})/2\times(60{,}000-70{,}000)=1.26\ m/s$$

$$vt= (6.20+1.26)\times 60/70=6.40\ m/s$$

When $v=46{,}500\ m/s,\ \alpha=26^o$ then $F_k= (1-1.39\times 10^{-4})F$ ($r_{sun}=1.20$ au), Oct 19, 2017

$$v_k - v=(-1.42\times 10^{-4}-1.39\times 10^{-4})/2\times(46{,}500-60{,}000)=1.90\ m/s$$

$$vt= (6.40+1.90)\times 46.5/60=6.43\ m/s$$

When $v=44{,}600\ m/s,\ \alpha=22^o$ then $F_k= (1-1.38\times 10^{-4})F$ ($r_{sun}=1.36$ au), Oct 25, 2017

$$v_k - v=(-1.39\times 10^{-4}-1.38\times 10^{-4})/2\times(44{,}600-46{,}500)=0.26\ m/s$$

$$vt=(6.43+0.26)\times 44.6/46.5=6.42\ m/s$$

On Oct 25, 2017 12:00 AM, CSS orbit viewer show the distance between Oumuamua and the Sun is 1.36 au (approximately 1.4 au),[19] the comparable acceleration $A_c$ should be:

$$A_{c1.36}=4.96\times 10^{-6}/1.36^2=2.68\times 10^{-6}\ m/s^2$$

At 1.4 au, $A_c$ should be:

$$A_{c1.40}=4.96\times 10^{-6}/1.40^2= 2.53\times 10^{-6}\ m/s^2$$

When $v=41{,}200\ m/s,\ \alpha=19^o$ then $F_k= (1-1.30\times 10^{-4})F$ ($r_{sun}=1.77$ au), Nov 12, 2017

$$v_k - v=(-1.38\times 10^{-4}-1.30\times 10^{-4})/2\times(41{,}200-44{,}600)=0.456\ m/s$$

$$vt=(6.42+0.456)\times 41.2/44.6=6.35\ m/s$$

From Oct 19, 2017, until the Hubble Space Telescope (HST) observation day on Nov 12, 2017, compare to the predicted position solely by gravity, the Oumuamua's trajectory deviation caused by the kinematic gravitational effect which expressed as $s_k - s$. In case without considering Oumuamua's velocity, during this period, assume $v_0 = 6.43$ *m/s*, $v_t = 6.35$ *m/s*, by using the equation $\bar{v} = (v0 + vt)/2$, $s = \bar{v}t$, $s_k - s$ can be calculated to be:

$s_k - s = (6.43+6.35)/2\times24\times86400 = 13{,}250{,}304m = 13{,}250\ km$

*When* $v = 37{,}700\ m/s$, $\alpha = 14^o$ *then* $F_k = (1-1.22\times10^{-4})F$ $(r_{sun} = 2.42\ au)$, *Dec 12, 2017*

$v_k - v = (-1.30\times10^{-4}-1.22\times10^{-4})/2\times(37{,}700-41{,}200) = 0.44\ m/s$

$vt = (6.35+0.44)\times37.7/41.2 = 6.21\ m/s$

$s_k - s = (6.35+6.21)/2\times30\times86400 = 16{,}277{,}760 = 16{,}278\ km$

Until the HST observation day on Dec 12, 2017, the Oumuamua's trajectory deviation caused by the kinematic gravitational effect can be calculated to be:

$13{,}250+16{,}278 = 29{,}528\ km$

*When* $v = 36{,}200\ m/s$, $\alpha = 12^o$ *then* $F_k = (1-1.18\times10^{-4})F$ $(r_{sun} = 2.86\ au)$, *Jan 2, 2018*

$v_k - v = (-1.22\times10^{-4}-1.18\times10^{-4})/2\times(36{,}200-37{,}700) = 0.18\ m/s$

$vt = (6.21+0.18)\times36.2/37.7 = 6.14\ m/s$

$s_k - s = (6.21+6.14)/2\times21\times86400 = 11{,}203{,}920m = 11{,}204\ km$

From Oct 19, 2017, until the HST observation day on Jan 2, 2018, compare with the predicted position solely by gravity, the Oumuamua's trajectory deviation caused by the kinematic gravitational effect of the Sun can be calculated to be:

$29{,}528+11{,}204 = 40{,}732\ km$

The Margin of error between the result calculated by KGE and the astronomical observation results is:

$(40{,}732-40{,}000)/40{,}000 \approx 0.018$

*When* $v = 32{,}200\ m/s$, $\alpha = 8^o$ *then* $F_k = (1-1.06\times10^{-4})F$ *May 3, 2018*

$v_k - v = (-1.18\times10^{-4}-1.06\times10^{-4}))/2\times(32{,}200-36{,}200) = 0.45\ m/s$

The velocity increment on May 3, 2018 should be:

$vt = (6.14+0.45)\times32.2/36.2 = 5.86\ m/s$

After the observation periods and until Oumuamua arrived at Jupiter's orbit, during *Jan 2, 2018 - May 3, 2018,* the its trajectory deviation between result of equation (1) and result of equation (2) can be calculated to be:

$s_k$ - $s$=(6.14+5.86)/2×121×86400= 62,726,400m=62,726 km

From Oct 19,2017 to May 3, 2018, it should be:

*40,732+62,726=103,458 km*

Additionally, the precision and time span of the trajectory simulations, the measurement accuracy of the α angle, and the precision of values used in the calculation process all influence the accuracy of the computed results.

**Specific calculation process of 3I/ATLAS's kinematic gravitational acceleration, accumulated velocity increments and trajectory deviations caused by KGE**

---

*When* $v$=58.1 km/s   $\alpha$=179°   $F_k$ = (1+1.94×10^{-4})F   $r_{Sun}$=133.56 au   Jan 1, 2015

Based on the acceleration due to gravity formula: $g=GM/r^2$, by substituting the gravitational constant $G=6.673\times10^{-11}$ $N.m^2/kg^2$ and the mass of the Sun $1.988\times10^{30}$ kg, and $1\ AU=1.496\times10^{11}$ m into the formula, the gravitational acceleration $a_g$ due to the Sun can be calculated.

$a_{g1.0}$=0.00593 m/s² = 2.96×10^{-4} au/day²

$a_{g133.56}$ = $a_{g1.0}$/133.56² = 3.32×10^{-7} m/s² = 1.66×10^{-8} au/day²

Before perihelion the instantaneous kinematic gravitational acceleration $a_k$ can be calculated by $(Fk - F)/F \times ag$ ; after perihelion $F$ will decelerate 3I/ATLAS and $F_k$ will less than $F$, decreasing deceleration equals a small acceleration, so $a_k$ can be calculated by $-(Fk - F)/F \times ag$ .

$a_k$ = 1.94×10^{-4}×2.96×10^{-4} /133.56² = 3.22×10^{-12} au/day²

*When* $v$=58.8 km/s   $\alpha$=175°   $F_k$ = (1+1.95×10^{-4})F   $r_{Sun}$=17.90 au   Jun 1, 2024

$v_k$ - $v$ = (1.94×10^{-4}+1.95×10^{-4})/2× (58,800-58,100)=0.14 m/s

$a_k$ = 1.95×10^{-4}×2.96×10^{-4} /17.90² = 1.80×10^{-10} au/day²

$A_{kc}$ = 0.14/86400 = 1.62×10^{-6} m/s^{-2} = 8.09×10^{-8} AU/day² = 12.10 km/day²

*When* $v$=60 km/s   $\alpha$=168°   $F_k$ = (1+1.96×10^{-4})F   $r_{Sun}$=7.28 au   Apr 10, 2025

Interval $v_k$ - $v$ = (1.95×10^{-4}+1.96×10^{-4})/2×(60,000-58,800)=0.23 m/s

Cumulative $vt$=0.14+0.23=0.37 m/s

$a_k$ = 1.96×10^{-4}×2.96×10^{-4} /7.28² = 1.09×10^{-9} au/day²

$A_{kc}$ = 0.37/86400 = 4.28×10^{-6} m/s^{-2} = 2.14×10^{-7} AU/day² = 31.97 km/day²

When $v$=61.3 km/s $\alpha$=161° $F_k = (1+1.93\times10^{-4})F$ $r_{Sun}$=4.52 au Jul 1, 2025

$v_k - v=(1.96\times10^{-4}+1.93\times10^{-4})/2\times(61,300-60,000)=0.25$ m/s

$vt=0.37+0.25=0.62$ m/s

$a_k=1.93\times10^{-4}\times2.96\times10^{-4}/4.52^2= 2.80\times10^{-9}$ au/day$^2$

$A_{kc}=0.62/86400=7.18\times10^{-6}$ m/s$^2$ $=3.58\times10^{-7}$ AU/day$^2$ =53.57 km/day$^2$

When $v$=62.5 km/s $\alpha$=152° $F_k = (1+1.84\times10^{-4})F$ $r_{Sun}$=3.23 au Aug 9, 2025

$v_k - v=(1.93\times10^{-4}+1.84\times10^{-4})/2\times(62,500-61,300)=0.23$ m/s

$vt=0.62+0.23=0.85$ m/s

$s_k - s= (0.62+0.85)/2\times39\times86400=2,476,656$m=2,477 km

$a_k=1.84\times10^{-4}\times2.96\times10^{-4}/3.23^2 =5.22\times10^{-9}$ au/day$^2$

$A_{kc}=0.85/86400=9.84\times10^{-6}$ m/s$^2$ $=4.91\times10^{-7}$ AU/day$^2$ =73.44 km/day$^2$

When $v$=64.1 km/s $\alpha$=143° $F_k =(1+1.71\times10^{-4})F$ $r_{Sun}$=2.40 au Sep 5, 2025

$v_k - v=(1.84\times10^{-4}+1.71\times10^{-4}))/2\times(64,100-62,500)=0.28$ m/s

$vt=0.85+0.28=1.13$ m/s

$s_k - s=(0.85+1.13)/2\times27\times86400=2,309,472$m=2,309 km

2,477+2,309=4,786 km

$a_k=1.71\times10^{-4}\times2.96\times10^{-4}/2.40^2=8.79\times10^{-9}$ au/day$^2$

$A_{kc}=1.13/86400=1.31\times10^{-5}$ m/s$^2$=$6.53\times10^{-7}$ AU/day$^2$=97.63 km/day$^2$

When $v$=65.7 km/s $\alpha$=131° $F_k = (1+1.44\times10^{-4})F$ $r_{Sun}$=1.85 au Sep 25, 2025

$v_k - v=(1.71\times10^{-4}+1.44\times10^{-4})/2\times(65,700-64,100)=0.25$ m/s

$vt=1.13+0.25=1.38$ m/s

$s_k - s= (1.13+1.38)/2\times20\times86400=2,168,640$m=2,169 km

4,786+2,169=6,955 km

$a_k=1.44\times10^{-4}\times2.96\times10^{-4}/1.85^2=1.25\times10^{-8}$ au/day$^2$

$A_{kc}=1.38/86400=1.60\times10^{-5}$ m/s$^2$=$7.97\times10^{-7}$ AU/day$^2$=119.23 km/day$^2$

When $v$=67.5 km/s $\alpha$=112° $F_k = (1+8.44\times10^{-5})F$ $r_{Sun}$=1.49 au Oct 13, 2025

$v_k - v=(1.44\times10^{-4}+8.44\times10^{-5})/2\times(67,500-65,700)=0.21$ m/s

$vt=1.38+0.21=1.59$ m/s

$s_k - s=(1.38+1.59)/2\times18\times86400=2,309,472$m=2,309 km

6,955+2,309=9,264 km

$a_k$=8.44×10$^{-5}$×2.96×10$^{-4}$/1.49$^2$=1.13×10$^{-8}$ au/day$^2$

$A_{kc}$=1.59/86400=1.84×10$^{-5}$ m/s$^2$ =9.18×10$^{-7}$ AU/day$^2$ =137.38 km/day$^2$

*When v=68.3 km/s* $\alpha$=90$^o$ $F_k$ = (1+2.60×10$^{-8}$)F $r_{Sun}$=1.36 au *Oct 29, 2025*

$v_k$ - v=(8.44×10$^{-5}$+2.60×10$^{-8}$)/2× (68,300-67,500) =0.03 m/s

*vt=1.59+0.03=1.62 m/s* *Perihelion*

For comparison the kinematic gravitational acceleration with the non-gravitational acceleration report,[29] theoretically, kinematic gravitational acceleration occurs throughout its entire solar system journey, but non-gravitational acceleration calculation assuming all previous velocity increments occurred within one day after passing perihelion. When 3I/ATLAS passing perihelion, the accumulated velocity increment of *1.62 m/s* was accumulated by kinematic gravitational acceleration during the period of Jan 1 2015 to Oct 29 2025 will lead to comparable accelerations $A_c$ and $A_{kc}$ at moving direction. Based on the calculation method of non-gravitational acceleration and acceleration formula: $\boldsymbol{a} = \frac{\boldsymbol{vt - v0}}{\boldsymbol{t}}$ , when $v_0$=0, $v_t$=1.62 m/s, *t=24×3600=86400s,* the acceleration should be:

$A_{c1.36}$=$A_{kc1.36}$=1.62/86400=1.875×10$^{-5}$ m/s$^{-2}$ =9.36×10$^{-7}$ AU/day$^2$ =139.97 km/day$^2$

$A_{c1}$= $A_{c\,1.36}$×1.36$^2$=3.47×10$^{-5}$ m/s$^{-2}$=1.73×10$^{-6}$ AU/day$^2$ = 258.89 km/day$^2$

$A_{1.0} = \sqrt{A1^2 + A2^2} = \sqrt{(1.66 \times 10^{-6})^2 + (7.09 \times 10^{-7})^2}$=3.62×10$^{-5}$ m/s$^2$=1.81×10$^{-6}$ AU/day$^2$ = 270.04 km/day$^{-2}$

$s_k$ - s=(1.59+1.62)/2×16×86400=2,218,752m=2,219 km

*9,264+2,219=11,483 km*

$a_k$=2.60×10$^{-8}$×2.96×10$^{-4}$/1.36$^2$=4.16×10$^{-12}$ au/day$^2$

*When v=67.5 km/s* $\alpha$=68$^o$ $F_k$ = (1-8.43×10$^{-5}$)F $r_{Sun}$=1.47 au *Nov 14, 2025*

$v_k$ - v=(2.60×10$^{-8}$-8.43×10$^{-5}$)/2×(67,500-68,300)=0.03 m/s

*vt=(1.62+0.03)×67.5/68.3=1.63 m/s*

$s_k$-s=(1.62+1.63)/2×16×86400=2,246,400m=2,246 km

*11,483+2,246=13,729 km*

$a_k$=8.43×10$^{-5}$×2.96×10$^{-4}$/1.47$^2$=1.15×10$^{-8}$ au/day$^2$

*When v=66 km/s* $\alpha$=52$^o$ $F_k$ = (1-1.36×10$^{-4}$)F $r_{Sun}$=1.78 au *Nov 30, 2025*

$v_k - v=(-8.43\times10^{-5}-1.36\times10^{-4})/2\times(66,000-67,500)=0.17\ m/s$

$vt=(1.63+0.17)\times66/67.5=1.76\ m/s$

$s_k - s= (1.63+1.76)/2\times16\times86400=2,343,168m=2,343\ km$

$13,729+2,343=16,072\ km$

$a_k=1.36\times10^{-4}\times2.96\times10^{-4}/1.78^2=1.27\times10^{-8}\ au/day^2$

*When* $v=64.3\ km/s$ $\alpha=39^o$ $F_k = (1-1.67\times10^{-4})F$ $r_{Sun}=2.30au$ *Dec 19, 2025*

$v_k - v=(-1.36\times10^{-4}-1.67\times10^{-4})/2\times(64,300-66,000)=0.26\ m/s$

$vt=(1.76+0.26)\times64.3/66=1.97\ m/s$

$s_k - s=(1.76+1.97)/2\times19\times86400=3,061,584m=3,062\ km$

When 3I/ATLAS arriving at the position nearest the Earth on Dec 19, 2025, relative to the predicted trajectory position solely by gravity, the kinematic gravitational acceleration of the Sun will make it travel an additional distance in its trajectory from Jul 1, 2025 to Dec 19, 2025, which should be:

$s_k - s=16,072+3,062=19,134\ km$

$a_k=1.67\times10^{-4}\times 2.96\times10^{-4}/2.30^2=9.34\times10^{-9}\ au/day^2$

*When* $v=63\ km/s$ $\alpha=30^o$ $F_k = (1-1.82\times10^{-4})F$ $r_{Sun}=2.95au$ *Jan 10, 2026*

$v_k - v=(-1.67\times10^{-4}-1.82\times10^{-4})/2\times(63,000-64,300)=0.23\ m/s$

$vt=(1.97+0.23)\times63/64.3=2.16\ m/s$

$s_k - s= (1.97+2.16)/2\times22\times86400=3,925,152m=3,925\ km$

$19,134+3,925=23,059\ km$

$a_k=1.82\times10^{-4}\times2.96\times10^{-4}/2.95^2=6.19\times10^{-9}\ au/day^2$

*When* $v=62\ km/s$ $\alpha=24^o$ $F_k = (1-1.89\times10^{-4})F$ $r_{Sun}=3.72au$ *Feb 3, 2026*

$v_k - v=(-1.82\times10^{-4}-1.89\times10^{-4})/2\times(62,000-63,000)=0.19\ m/s$

$vt=(2.16+0.19)\times62/63=2.31\ m/s$

$s_k - s= (2.16+2.31)/2\times24\times86400=4,634,496m=4,634\ km$

$23,059+4,634=27,693\ km$

$a_k=1.89\times10^{-4}\times2.96\times10^{-4}/3.72^2=4.04\times10^{-9}\ au/day^2$

*When* $v=61.3\ km/s$ $\alpha=19^o$ $F_k=(1-1.93\times10^{-4})F$ $r_{Sun}=4.45au$ *Feb 25, 2026*

$v_k - v=(-1.89\times10^{-4}-1.93\times10^{-4})/2\times(61,300-62,000)=0.13\ m/s$

$vt=(2.31+0.13)\times61.3/62=2.41\ m/s$

$s_k$ - $s= (2.31+2.41)/2\times22\times86400=4,485,888m=4,486\ km$

$27,693+4,486=32,179\ km$

$a_k=1.93\times10^{-4}\times2.96\times10^{-4}/4.45^2=2.88\times10^{-9}\ au/day^2$

*When* $v=60.8\ km/s$ $\alpha=16^o$ $F_k=(1-1.95\times10^{-4})F$ $r_{Sun}=5.22au$ *Mar 20, 2026*

$v_k$ - $v=(-1.93\times10^{-4}-1.95\times10^{-4})/2\times(60,800-61,300)=0.10\ m/s$

$vt=(2.41+0.10)\times60.8/61.3=2.49\ m/s$

$s_k$ - $s= (2.41+2.49)/2\times23\times86400=4,868,640m=4,869\ km$

When 3I/Atlas near Jupiter's orbit on Mar 20, 2026*,* during the period of Jul 1 2025 – Mar 20 2026, relative to the predicted trajectory position solely by gravity, the kinematic gravitational acceleration of the Sun will make it travel an additional distance in its trajectory, which should be:

$s_k$ - $s=32,179+4,869=37,048\ km$

$a_k=1.95\times10^{-4}\times2.96\times10^{-4}/5.22^2=2.12\times10^{-9}\ au/day^2$

**Specific calculation process of the accumulated velocity increments of C/2025 R3 (Pan-STARRS) caused by KGE**

---

*When* $v=1.8\ km/s$ $\alpha=179^o$ $F_k=(1+6.00\times10^{-6})F$ $r=571.46au$ *Jan 01, 1000*

*When* $v=10.0\ km/s$ $\alpha=172^o$ $Fk=(1+3.30\times10-5)F$ $r=17.92au$ *May 18, 2020*

$v_k$ - $v=(6.00\times10^{-6}+3.30\times10^{-5})/2\times(10,000-1,800)=0.16\ m/s$

$A_{kc}=0.16/86400=1.85\times10^{-6}\ m/s^2=9.24\times10^{-8}\ AU/day^2=13.82\ km/day^2$

*When* $v=20.0\ km/s$ $\alpha=159^o$ $F_k=(1+6.23\times10^{-5})F$ $r=4.45\ au$ *Jun 28, 2025*

*Interval* $v_k$ - $v=(3.30\times10^{-5}+6.23\times10^{-5})/2\times(20,000-10,000)=0.48\ m/s$

*Cumulative* $vt=0.16+0.48=0.64\ m/s$

$A_{kc}=0.64/86400=7.41\times10^{-6}\ m/s^2=3.70\times10^{-7}\ AU/day^2=55.30\ km/day^2$

*When* $v=30\ km/s$ $\alpha=149^o$ $F_k = (1+8.58\times10^{-5})F$ $r=1.97au$ *Jan 12, 2026*

$v_k$ - $v=(6.23\times10^{-5}+8.58\times10^{-5})/2\times(30,000-20,000)= 0.74m/s$

$vt=0.64+0.74=1.38\ m/s$

$A_{kc}=1.38/86400=1.60\times10^{-5}\ m/s^2=7.97\times10^{-7}\ AU/day^2=119.23\ km/day^2$

*When* $v=39.8\ km/s$ $\alpha=138^o$ $F_k = (1+9.87\times10^{-5})F$ $r=1.12au$ *Mar 06, 2026*

$v_k$ - $v=(8.58\times10^{-5}+9.87\times10^{-5})/2\times(39,800-30,000)=0.90\ m/s$

$vt=1.38+0.90=2.28\ m/s$

$A_{kc}=2.28/86400=2.64\times10^{-5}\ m/s^2=1.32\times10^{-6}\ AU/day^2=196.99\ km/day^2$

When $v=45.8\ km/s$ $\alpha=133^o$ $F_k=(1+1.04\times10^{-4})F$ $r=0.85au$ Mar 22, 2026

$v_k-v=(9.87\times10^{-5}+1.04\times10^{-4})/2\times(45{,}800-39{,}800)=0.61\ m/s$

$vt=2.28+0.61=2.89\ m/s$

$A_{kc}=2.89/86400=3.34\times10^{-5}\ m/s^2=1.67\times10^{-6}\ AU/day^2=249.70\ km/day^2$

When $v=51.0\ km/s$ $\alpha=126^o$ $F_k=(1+1.00\times10^{-4})F$ $r=0.68au$ Apr 01, 2026

$v_k-v=(1.04\times10^{-4}+1.00\times10^{-4})/2\times(51{,}000-45{,}800)=0.53\ m/s$

$vt=2.89+0.53=3.42\ m/s$

$A_{kc}=3.42/86400=3.96\times10^{-5}\ m/s^2=1.98\times10^{-6}\ AU/day^2=295.49\ km/day^2$

When $v=56.2\ km/s$ $\alpha=115^o$ $F_k=(1+7.92\times10^{-5})F$ $r=0.56au$ Apr 10, 2026

$v_k-v=(1.00\times10^{-4}+7.92\times10^{-5})/2\times(56{,}200-51{,}000)=0.47\ m/s$

$vt=3.42+0.47=3.89\ m/s$

$A_{kc}=3.89/86400=4.50\times10^{-5}\ m/s^2=2.25\times10^{-6}\ AU/day^2=336.10\ km/day^2$

When $v=59.7\ km/s$ $\alpha=90^o$ $F_k=(1+1.98\times10^{-8})F$ $r=0.499au$ Apr 20, 2026

$v_k-v=(7.92\times10^{-5}+1.98\times10^{-8})/2\times(59{,}700-56{,}200)=0.14\ m/s$

$vt=3.89+0.14=4.03\ m/s$ Perihelion

$A_{kc}=4.03/86400=4.66\times10^{-5}\ m/s^2=2.33\times10^{-6}\ AU/day^2=348.20\ km/day^2$

When $v=56.5\ km/s$ $\alpha=67^o$ $F_k=(1-7.36\times10^{-5})F$ $r=0.56au$ May 01, 2026

$v_k-v=(1.98\times10^{-8}-7.36\times10^{-5})/2\times(56{,}500-59{,}700)=0.12$

$vt=(4.03+0.12)\times56.5/59.7=3.93\ m/s$

When $v=51.3\ km/s$ $\alpha=56^o$ $F_k=(1-9.57\times10^{-5})F$ $r=0.67au$ May 10, 2026

$v_k-v=(-7.36\times10^{-5}-9.57\times10^{-5})/2\times(51{,}300-56{,}500)=0.44\ m/s$

$vt=(3.93+0.44)\times51.3/56.5=3.97\ m/s$

When $v=46.0\ km/s$ $\alpha=48^o$ $F_k=(1-1.03\times10^{-4})F$ $r=0.84au$ May 20, 2026

$v_k-v=(-9.57\times10^{-5}-1.03\times10^{-4})/2\times(46{,}000-51{,}300)=0.53\ m/s$

$vt=(3.97+0.53)\times46.0/51.3=4.04\ m/s$

When $v=40.0\ km/s$ $\alpha=43^o$ $F_k=(1-9.76\times10^{-5})F$ $r=1.11au$ Jun 05, 2026

$v_k-v=(-1.03\times10^{-4}-9.76\times10^{-5})/2\times(40{,}000-46{,}000)=0.60\ m/s$

$vt=(4.04+0.60)\times40.0/46.0=4.03\ m/s$

*When* $v=29.9$ *km/s* $\alpha=31^o$ $F_k=(1-8.55\times10^{-5})F$ $r=1.98au$ *Jul 29, 2026*

$v_k - v= (-9.76\times10^{-5}-8.55\times10^{-5})/2\times (29{,}900-40{,}000) =0.92$ *m/s*

$vt=(4.03+0.92)\times29.9/40.0=3.70$ *m/s*

*When* $v=20.0$ *km/s* $\alpha=21^o$ $F_k= (1-6.23\times10^{-5})F$ $r=4.42au$ *Feb 09, 2027*

$v_k - v= (-8.55\times10^{-5}-6.23\times10^{-5})/2\times (20{,}000-29{,}900) =0.73$ *m/s*

$vt=(3.70+0.73)\times20.0/29.9=2.96$ *m/s*

## Acknowledgment

CSS Orbit Viewer, CSS, D. Rankin, Spacein3d, Marco Micheli, Davide Farnocchia, Avi Loeb.

## Data availability

All data included in this study are available upon request by contact with the corresponding author.

## References

1. Jet Propulsion Laboratory. Our Solar System's First Known Interstellar Object Gets Unexpected Speed Boost (2018) https://www.jpl.nasa.gov/news/our-solar-systems-first-known-interstellar-object-gets-unexpected-speed-boost
2. The European Space Agency. Oumuamua's journey through our Solar System (2018) https://www.esa.int/ESA_Multimedia/Images/2018/06/Oumuamua_s_journey_through_our_Solar_System
3. Klesman, A. Oumuamua: our first interstellar visitor (2023) https://www.astronomy.com/science/oumuamua-our-first-interstellar-visitor
4. Micheli, M. *et al.* Non-gravitational acceleration in the trajectory of 1I/2017 U1 ('Oumuamua). *Nature* **559**, 223-226 (2018).
5. Seligman, D., Laughlin, G. & Batygin, K. On the Anomalous Acceleration of 1I/2017 U1 'Oumuamua. *The Astrophysical Journal Letters* (2019).

6. Yeomans, D. K. Cometary Orbital Dynamics and Astrometry. *International Astronomical Union Colloquium* (1991).
7. Gunnarsson, M., Bockelee-Morvan, D., Winnberg, A., Rickman, H. & Rantakyr, F. T. Production and kinematics of CO in comet C/1995O1(Hale-Bopp) at large post-perihelion distances. *Astronomy & Astrophysics* **402**, 383-393 (2003).
8. Weissman, P. & R. Nongravitational perturbations of long-period comets. *The Astronomical Journal* **84**, 580-580 (1979).
9. Maquet, L., Colas, F., Jorda, L. & Crovisier, J. CONGO, model of cometary non-gravitational forces combining astrometric and production rate data Application to comet 19P/Borrelly. *Astronomy & Astrophysics* **548** (2012).
10. Meech, K. J. *et al.* A brief visit from a red and extremely elongated interstellar asteroid. *Nature* **552**, 378-381 (2017).
11. Jewitt, D. *et al.* Interstellar Interloper 1I/2017 U1: Observations from the NOT and WIYN Telescopes. *The Astrophysical Journal Letters* **850**, L36 (2017).
12. Ye, Q. Z., Zhang, Q., Kelley, M. S. P. & Brown, P. G. 1I/2017 U1 ('Oumuamua) is Hot: Imaging, Spectroscopy, and Search of Meteor Activity. *Astrophysical Journal* **851** (2017).
13. Trilling, D. E., Mommert, M., Hora, J. L., Farnocchia, D. & Micheli, M. Spitzer Observations of Interstellar Object 1I/'Oumuamua. *The Astronomical Journal* **156**, 261 (2018).
14. Bialy, S. & Loeb, A. Could Solar Radiation Pressure Explain 'Oumuamua's Peculiar Acceleration? *The Astrophysical Journal Letters* **868**, L1 (2018).
15. Curran, S. J. 'Oumuamua as a light sail -- evidence against artificial origin. *Astronomy and Astrophysics* (2021).
16. Seligman, D. & Laughlin, G. Evidence that 1I/2017 U1 ('Oumuamua) was Composed of Molecular Hydrogen Ice. *The Astrophysical Journal Letters* **896**, L8 (2020).
17. Hoang, T. & Loeb, A. Destruction of Molecular Hydrogen Ice and Implications for

1I/2017 U1 ('Oumuamua). *The Astrophysical Journal Letters* **899**, L23 (2020).

18. Bergner, J.B. & Seligman, D. Z. Acceleration of 1I/'Oumuamua from radiolytically produced H2 in H2O ice. *Nature* **615**, 610-613 (2023).
19. Catalina Sky Survey Orbit Viewer https://neofixer.arizona.edu/css-orbit-view (2025)
20. International Astronomical Union Minor Planet Center https://minorplanetcenter.net/db_search/show_object?object_id=1I (2017)
21. International Astronomical Union Minor Planet Center https://minorplanetcenter.net/db_search/show_object?object_id=3I (2025)
22. Spacein3D. Where is interstellar asteroid ' Oumuamua? Live tracker https://spacein3d.com/asteroid-oumuamua-live-tracker (2023)
23. Feinstein, Adina D. et al. Precovery Observations of 3I/ATLAS from TESS Suggest Possible Distant Activity. *The Astrophysical Journal Letters* (2025)
24. Jewitt, David C. et al. Hubble Space Telescope Observations of the Interstellar Interloper 3I/ATLAS. *The Astrophysical Journal Letters* (2025)
25. Santana-Ros, Toni et al. Temporal evolution of the third interstellar comet 3I/ATLAS: Spin, color, spectra, and dust activity. *Astronomy & Astrophysics* (2025)
26. Seligman, Darryl Z. et al. Discovery and Preliminary Characterization of a Third Interstellar Object: 3I/ATLAS (2025) https://arxiv.org/html/2507.02757v3
27. Chandler, C.O., Bernardinelli, P.H., Jurić, M., et al., NSF-DOE Vera C. Rubin Observatory Observations of Interstellar Comet 3I/ATLAS (C/2025 N1) (2025) arXiv:2507.13409 https://arxiv.org/abs/2507.13409
28. Martinez-Palomera, J., Tuson, A., Hedges, C., et al., Pre-discovery TESS Observations of Interstellar Object 3I/ATLAS (2025) arXiv:2508.02499 https://arxiv.org/abs/2508.02499
29. Davide Farnocchia. C/2025 N1 (ATLAS) JPL 28, Epoch 2460882.5 (update Oct 29 2025)

https://ssd.jpl.nasa.gov/tools/sbdb_lookup.html#/?sstr=3I%2FATLAS (displayed in Supplementary Materials 1)

30. Avi-Loeb. First-Evidence-for-a-Non-Gravitational-Acceleration-of-3i-atlas-at-Perihelion (2025) https://avi-loeb.medium.com/first-evidence-for-a-non-gravitational-acceleration-of-3i-atlas-at-perihelion-2698f6a453fe
31. Avi-Loeb. Afterthoughts on the Non-Gravitational Acceleration of 3I/ATLAS at Perihelion (2025) https://avi-loeb.medium.com/afterthoughts-on-the-non-gravitational-acceleration-of-3i-atlas-at-perihelion-97c609e24fc6
32. Dongcheng Zhao. Using Modified Newtonian Dynamics to Explain the Anomalous Acceleration of Oumuamua and Incompletely Predict the Additional Acceleration of 3I/ATLAS (Oct 27 2025) (Partial version) https://arxiv.org/abs/2010.00228v14
33. Dongcheng Zhao. Using Modified Newtonian Dynamics to Explain the Anomalous Acceleration of Oumuamua and Incompletely Predict the Additional Acceleration of 3I/ATLAS (Oct 27 2025) (Full version) https://storage.googleapis.com/sn-mts-ejp-nature/nature_files/2025/10/27/00535387/00/535387_0_art_file_4931831_t4s4wt.docx

## Supplementary Materials

---

1. Observed additional acceleration of 3I/ATLAS released by Davide Farnocchia on Oct 29, 2025.

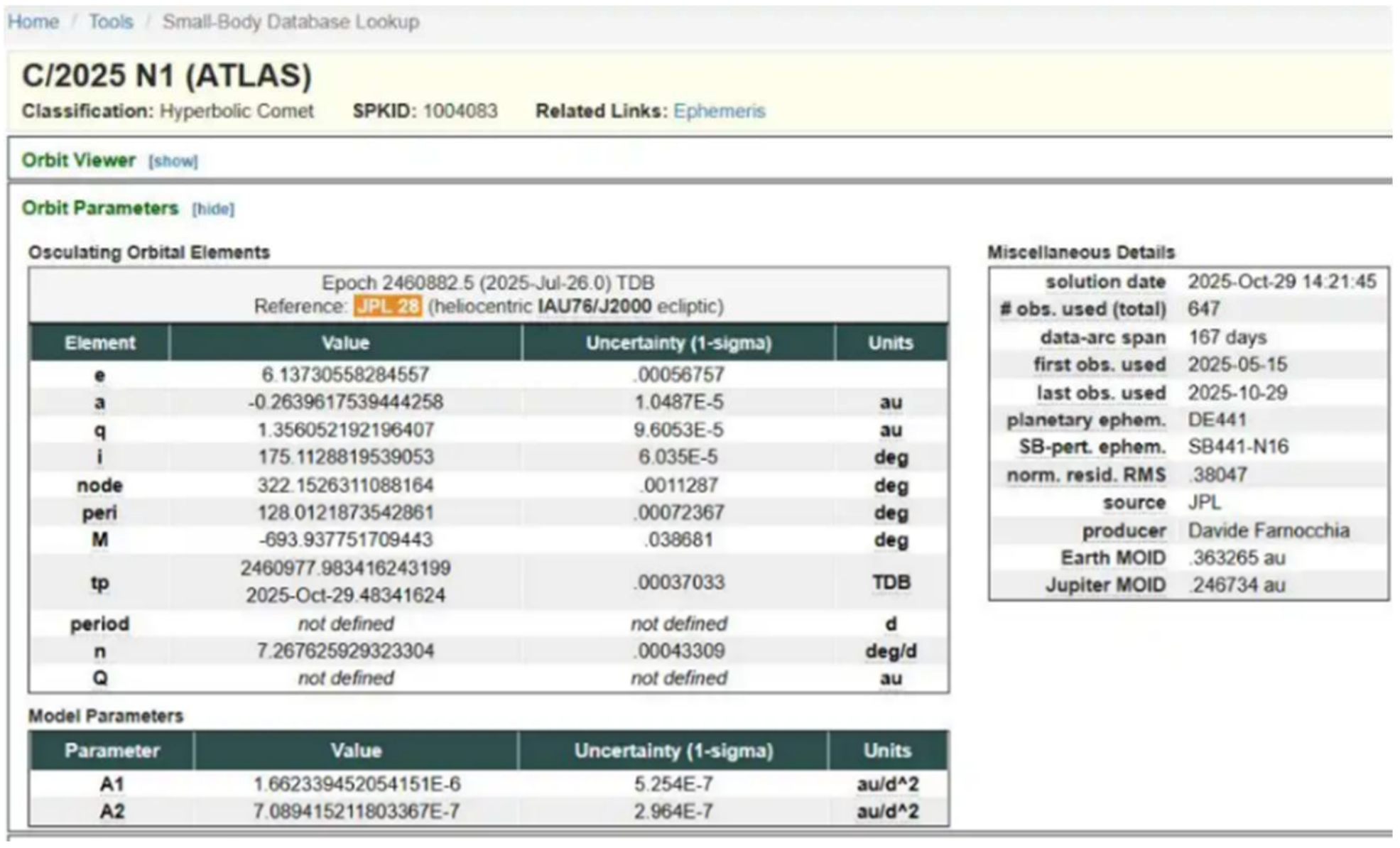

Home / Tools / Small-Body Database Lookup

**C/2025 N1 (ATLAS)**

**Classification:** Hyperbolic Comet **SPKID:** 1004083 **Related Links:** Ephemeris

**Orbit Viewer** [show]

**Orbit Parameters** [hide]

**Osculating Orbital Elements**

Epoch 2460882.5 (2025-Jul-26.0) TDB
Reference: JPL 28 (heliocentric **IAU76/J2000** ecliptic)

| Element | Value | Uncertainty (1-sigma) | Units |
|---|---|---|---|
| e | 6.13730558284557 | .00056757 | |
| a | -0.2639617539444258 | 1.0487E-5 | au |
| q | 1.356052192196407 | 9.6053E-5 | au |
| i | 175.1128819539053 | 6.035E-5 | deg |
| node | 322.1526311088164 | .0011287 | deg |
| peri | 128.0121873542861 | .00072367 | deg |
| M | -693.937751709443 | .038681 | deg |
| tp | 2460977.983416243199<br>2025-Oct-29.48341624 | .00037033 | TDB |
| period | *not defined* | *not defined* | d |
| n | 7.267625929323304 | .00043309 | deg/d |
| Q | *not defined* | *not defined* | au |

**Model Parameters**

| Parameter | Value | Uncertainty (1-sigma) | Units |
|---|---|---|---|
| A1 | 1.662339452054151E-6 | 5.254E-7 | au/d^2 |
| A2 | 7.089415211803367E-7 | 2.964E-7 | au/d^2 |

**Miscellaneous Details**

| | |
|---|---|
| solution date | 2025-Oct-29 14:21:45 |
| # obs. used (total) | 647 |
| data-arc span | 167 days |
| first obs. used | 2025-05-15 |
| last obs. used | 2025-10-29 |
| planetary ephem. | DE441 |
| SB-pert. ephem. | SB441-N16 |
| norm. resid. RMS | .38047 |
| source | JPL |
| producer | Davide Farnocchia |
| Earth MOID | .363265 au |
| Jupiter MOID | .246734 au |

2. Non-gravitational acceleration of 3I/ATLAS for $co_2$ released by Davide Farnocchia on Jan 08, 2026.

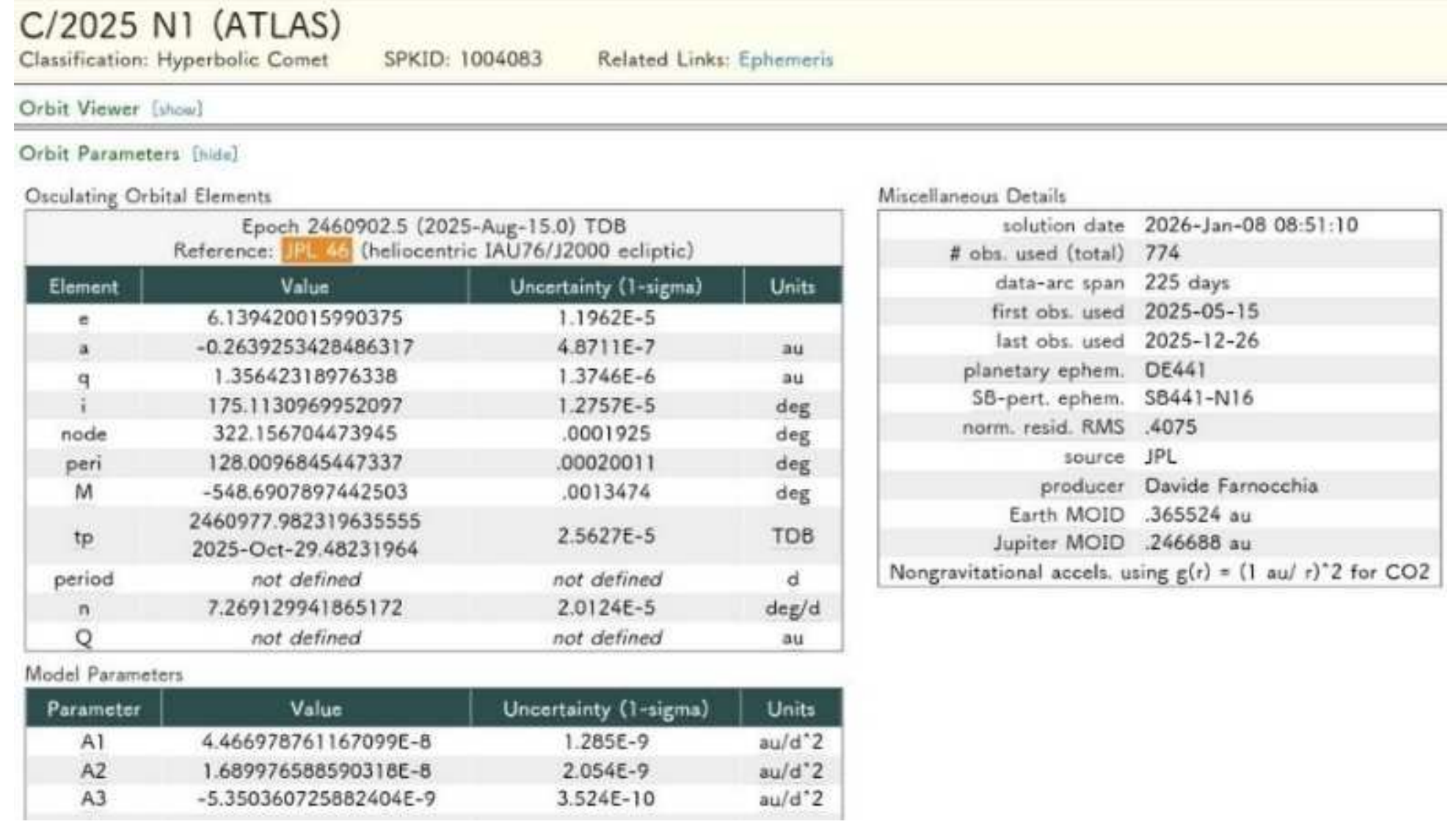

C/2025 N1 (ATLAS)

Classification: Hyperbolic Comet SPKID: 1004083 Related Links: Ephemeris

Orbit Viewer [show]

Orbit Parameters [hide]

Osculating Orbital Elements

Epoch 2460902.5 (2025-Aug-15.0) TDB
Reference: JPL 46 (heliocentric IAU76/J2000 ecliptic)

| Element | Value | Uncertainty (1-sigma) | Units |
|---|---|---|---|
| e | 6.139420015990375 | 1.1962E-5 | |
| a | -0.2639253428486317 | 4.8711E-7 | au |
| q | 1.35642318976338 | 1.3746E-6 | au |
| i | 175.1130969952097 | 1.2757E-5 | deg |
| node | 322.156704473945 | .0001925 | deg |
| peri | 128.0096845447337 | .00020011 | deg |
| M | -548.6907897442503 | .0013474 | deg |
| tp | 2460977.982319635555<br>2025-Oct-29.48231964 | 2.5627E-5 | TDB |
| period | *not defined* | *not defined* | d |
| n | 7.269129941865172 | 2.0124E-5 | deg/d |
| Q | *not defined* | *not defined* | au |

Model Parameters

| Parameter | Value | Uncertainty (1-sigma) | Units |
|---|---|---|---|
| A1 | 4.466978761167099E-8 | 1.285E-9 | au/d^2 |
| A2 | 1.689976588590318E-8 | 2.054E-9 | au/d^2 |
| A3 | -5.350360725882404E-9 | 3.524E-10 | au/d^2 |

Miscellaneous Details

| | |
|---|---|
| solution date | 2026-Jan-08 08:51:10 |
| # obs. used (total) | 774 |
| data-arc span | 225 days |
| first obs. used | 2025-05-15 |
| last obs. used | 2025-12-26 |
| planetary ephem. | DE441 |
| SB-pert. ephem. | SB441-N16 |
| norm. resid. RMS | .4075 |
| source | JPL |
| producer | Davide Farnocchia |
| Earth MOID | .365524 au |
| Jupiter MOID | .246688 au |
| Nongravitational accels. using g(r) = (1 au/ r)^2 for CO2 | |

3. Non-gravitational acceleration of 3I/ATLAS for $co_2$ released by Davide Farnocchia on Jan 13, 2026.

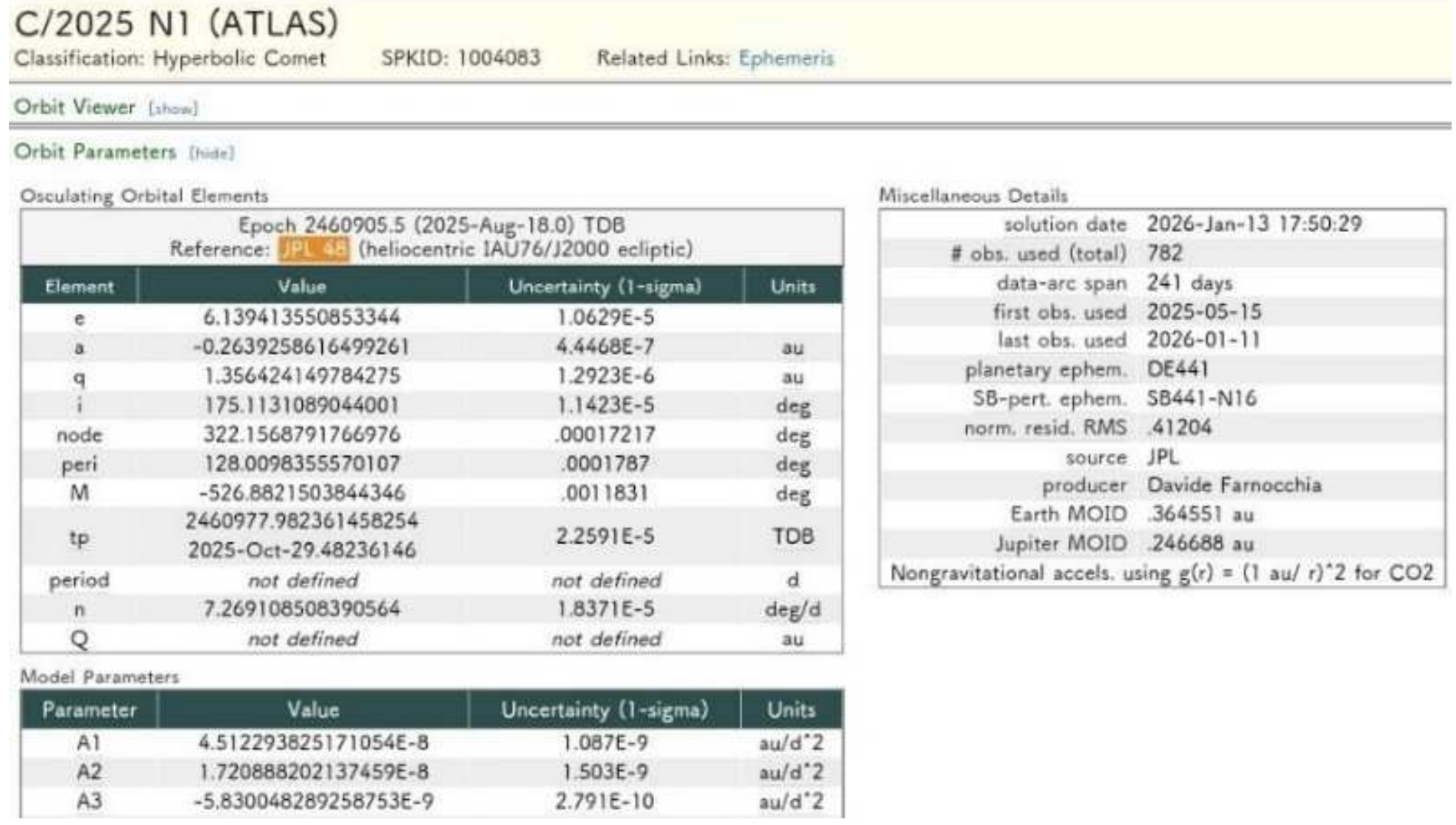

## C/2025 N1 (ATLAS)

Classification: Hyperbolic Comet    SPKID: 1004083    Related Links: Ephemeris

Orbit Viewer [show]

Orbit Parameters [hide]

Osculating Orbital Elements

Epoch 2460905.5 (2025-Aug-18.0) TDB
Reference: JPL 48 (heliocentric IAU76/J2000 ecliptic)

| Element | Value | Uncertainty (1-sigma) | Units |
|---|---|---|---|
| e | 6.139413550853344 | 1.0629E-5 | |
| a | -0.2639258616499261 | 4.4468E-7 | au |
| q | 1.356424149784275 | 1.2923E-6 | au |
| i | 175.1131089044001 | 1.1423E-5 | deg |
| node | 322.1568791766976 | .00017217 | deg |
| peri | 128.0098355570107 | .0001787 | deg |
| M | -526.8821503844346 | .0011831 | deg |
| tp | 2460977.982361458254<br>2025-Oct-29.48236146 | 2.2591E-5 | TDB |
| period | *not defined* | *not defined* | d |
| n | 7.269108508390564 | 1.8371E-5 | deg/d |
| Q | *not defined* | *not defined* | au |

Model Parameters

| Parameter | Value | Uncertainty (1-sigma) | Units |
|---|---|---|---|
| A1 | 4.512293825171054E-8 | 1.087E-9 | au/d^2 |
| A2 | 1.720888202137459E-8 | 1.503E-9 | au/d^2 |
| A3 | -5.830048289258753E-9 | 2.791E-10 | au/d^2 |

Miscellaneous Details

| | |
|---|---|
| solution date | 2026-Jan-13 17:50:29 |
| # obs. used (total) | 782 |
| data-arc span | 241 days |
| first obs. used | 2025-05-15 |
| last obs. used | 2026-01-11 |
| planetary ephem. | DE441 |
| SB-pert. ephem. | SB441-N16 |
| norm. resid. RMS | .41204 |
| source | JPL |
| producer | Davide Farnocchia |
| Earth MOID | .364551 au |
| Jupiter MOID | .246688 au |
| Nongravitational accels. using g(r) = (1 au/ r)^2 for CO2 | |

4. Non-gravitational acceleration of 3I/ATLAS for $co_2$ released by Davide Farnocchia on Jan 21, 2026.

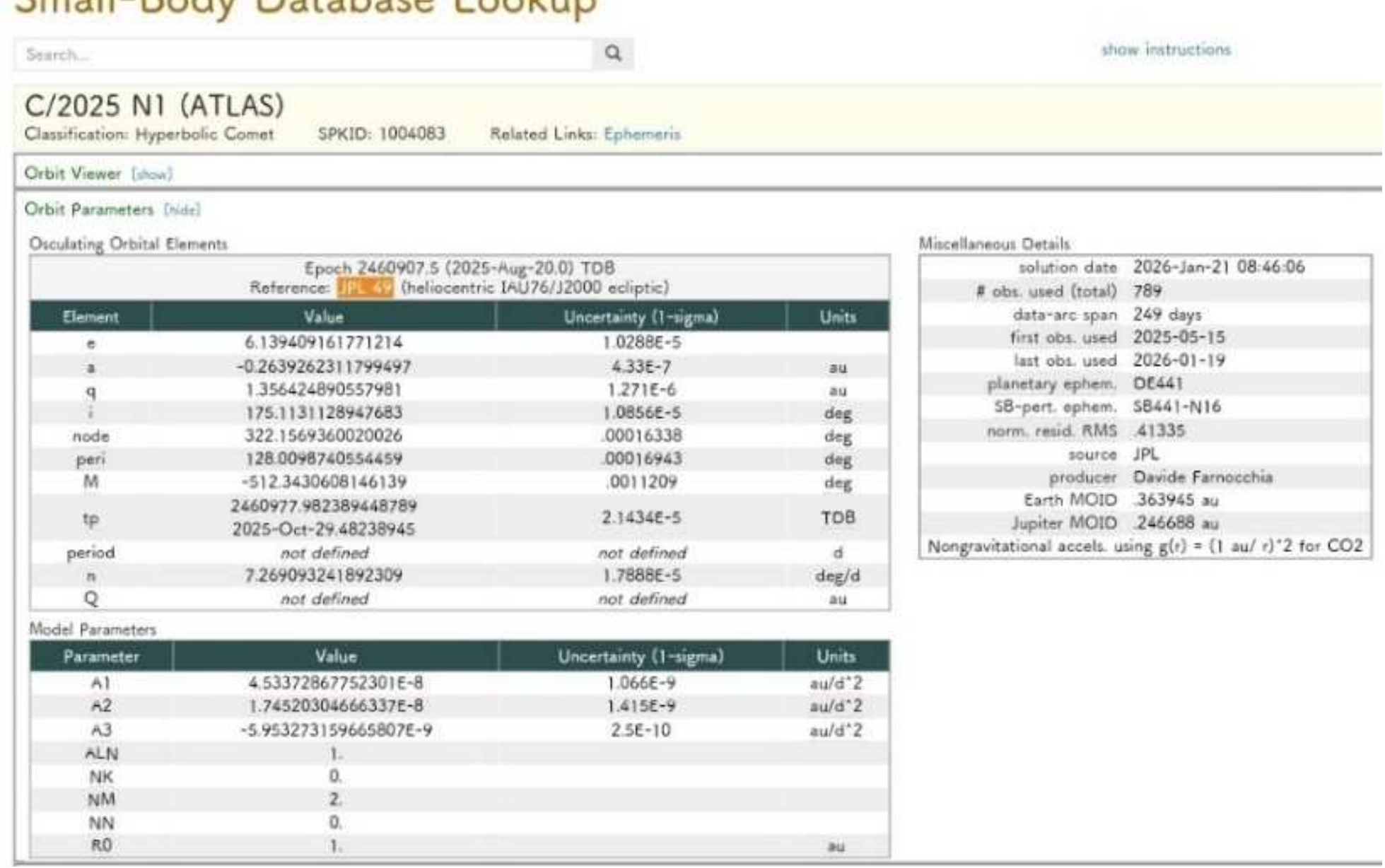

## Small-Body Database Lookup

Search...    show instructions

## C/2025 N1 (ATLAS)

Classification: Hyperbolic Comet    SPKID: 1004083    Related Links: Ephemeris

Orbit Viewer [show]

Orbit Parameters [hide]

Osculating Orbital Elements

Epoch 2460907.5 (2025-Aug-20.0) TDB
Reference: JPL 49 (heliocentric IAU76/J2000 ecliptic)

| Element | Value | Uncertainty (1-sigma) | Units |
|---|---|---|---|
| e | 6.139409161771214 | 1.0288E-5 | |
| a | -0.2639262311799497 | 4.33E-7 | au |
| q | 1.356424890557981 | 1.271E-6 | au |
| i | 175.1131128947683 | 1.0856E-5 | deg |
| node | 322.1569360020026 | .00016338 | deg |
| peri | 128.0098740554459 | .00016943 | deg |
| M | -512.3430608146139 | .0011209 | deg |
| tp | 2460977.982389448789<br>2025-Oct-29.48238945 | 2.1434E-5 | TDB |
| period | *not defined* | *not defined* | d |
| n | 7.269093241892309 | 1.7888E-5 | deg/d |
| Q | *not defined* | *not defined* | au |

Model Parameters

| Parameter | Value | Uncertainty (1-sigma) | Units |
|---|---|---|---|
| A1 | 4.53372867752301E-8 | 1.066E-9 | au/d^2 |
| A2 | 1.74520304666337E-8 | 1.415E-9 | au/d^2 |
| A3 | -5.953273159665807E-9 | 2.5E-10 | au/d^2 |
| ALN | 1. | | |
| NK | 0. | | |
| NM | 2. | | |
| NN | 0. | | |
| R0 | 1. | | au |

Miscellaneous Details

| | |
|---|---|
| solution date | 2026-Jan-21 08:46:06 |
| # obs. used (total) | 789 |
| data-arc span | 249 days |
| first obs. used | 2025-05-15 |
| last obs. used | 2026-01-19 |
| planetary ephem. | DE441 |
| SB-pert. ephem. | SB441-N16 |
| norm. resid. RMS | .41335 |
| source | JPL |
| producer | Davide Farnocchia |
| Earth MOID | .363945 au |
| Jupiter MOID | .246688 au |
| Nongravitational accels. using g(r) = (1 au/ r)^2 for CO2 | |